\documentclass[aip,
rsi,%
 amsmath,amssymb,
 reprint,%
]{revtex4-1}

\usepackage{graphicx}
\usepackage{dcolumn}
\usepackage{bm}
\usepackage[version=3]{mhchem}
\usepackage{threeparttable}
\usepackage{color}
\usepackage{xr}
\begin{document}

\title{Permutationally invariant polynomial regression for energies and gradients, using reverse differentiation, achieves orders of magnitude speed-up with high precision compared to other machine learning methods}
\date{\today}
\author{Paul L. Houston}
\email{plh2@cornell.edu}
\affiliation{Department of Chemistry and Chemical Biology, Cornell University, Ithaca, New York
14853, U.S.A. and Department of Chemistry and Biochemistry, Georgia Institute of
Technology, Atlanta, Georgia 30332, U.S.A}
\author{Chen Qu}
\affiliation{Department of Chemistry \& Biochemistry, University of Maryland, College Park, Maryland 20742, U.S.A.}
\author{Apurba Nandi}
\email{apurba.nandi@emory.edu}
\affiliation{Department of Chemistry and Cherry L. Emerson Center for Scientific Computation, Emory University, Atlanta, Georgia 30322, U.S.A.}
\author{Riccardo Conte}
\email{riccardo.conte1@unimi.it}
\affiliation{Dipartimento di Chimica, Universit\`{a} Degli Studi di Milano, via Golgi 19, 20133 Milano, Italy}
\author{Qi Yu}
\email{q.yu@yale.edu}
\affiliation{Department of Chemistry, Yale University, New Haven, Connecticut, U.S.A.}
\author{Joel M. Bowman}
\email{jmbowma@emory.edu}
\affiliation{Department of Chemistry and Cherry L. Emerson Center for Scientific Computation, Emory University, Atlanta, Georgia 30322, U.S.A.}

\begin{abstract}
Permutationally invariant polynomial (PIP) regression has been used to obtain machine-learned (ML) potential energy surfaces, including analytical gradients, for many molecules and chemical reactions. Recently, the approach has been extended to moderate size molecules and applied to systems up to 15 atoms. The algorithm, including ``purification of the basis'', is computationally efficient for energies; however, we found that the recent extension to obtain analytical gradients, despite being a remarkable advance over previous methods, could be further improved. Here we report developments to compact further a purified basis and, more significantly, to use the reverse gradient approach to greatly speed up gradient evaluation. We demonstrate this for our recent 4-body water interaction potential. Comparisons of training and testing precision on the MD17 database of energies and gradients (forces) for ethanol against GP-SOAP, ANI, sGDML, PhysNet, KREG, KRR and other methods, which were recently assessed by Dral and co-workers, are given. The PIP fits are as precise as those using these methods, but the PIP computation time for energy and force evaluation is shown to be 10 to 1000 times faster. Finally, a new PIP PES is reported for ethanol based on a more extensive dataset of energies and gradients than in the MD17 database. Diffusion Monte Carlo calculations which fail on MD17-based PESs are successful using the new PES.  
\end{abstract}
\maketitle

\section*{Introduction}
There has been dramatic progress in using regression methods from machine learning (ML) to develop high-dimensional potential energy surfaces (PESs).  This has also led to a plethora of perspectives in this field which are beyond the scope of this article to review. But we do note that in 2020-present there have been at least 8 perspectives in the mainline journals, J. Chem. Phys. and J. Phys. Chem. Letters.\cite{persp1,persp2,persp3,perps4,persp5,persp6,persp7,persp8} These excellent papers convey the breadth and excitement of this important application of ML to potentials. 

Perhaps the first Perspective in J. Phys. Chem. Lett. on this topic came from the Bowman group in 2010,\cite{persp9} where the theory and numerous applications of permutationally invariant polynomial (PIP) regression were introduced to the readers of this journal.  

This ML method, introduced for \ce{CH5+} in 2003,\cite{Bowman2003} is actively used\cite{ARPC2018,robo20,mbpoltests,paespips_21} and further developed.\cite{conte20_efficientPIP,jasperpips_21,greedypip_21} PIPs have also been incorporated in Neural Network methods,\cite{Guo16,persp6,NNZhang16,FINN_18}, Gaussian Process Regression\cite{PIP-GP}, and recently to atom-centered Gaussian Process Regression.\cite{Xie10Allen2021,Oord2020}

There are now numerous ML methods and codes to fit electronic energies and energies plus gradients (forces), and many of these are the subjects of the perspectives mentioned above as well as many more reviews and perspectives over the past 10 years.  It is clearly important to assess the performance of these various methods on the same datasets and ideally run on the same computer. 

This was recently done for ML methods applied to molecules with 9 or more atoms in several studies.\cite{paescomps, PIP-GP, Tkatchjcp, dralchemsci} The paper by Pinbeiro et al.\cite{dralchemsci} is particularly noteworthy as it contains a comprehensive study for ethanol, using the the MD17 dataset of energies and forces\cite{MD17}, of the dependence of precision and prediction times on training size for energies and energies plus forces for several popular approaches to constructing ML PESs, such as GAP-SOAP,\cite{GP-2015-1} ANI,\cite{AN1} DPMD,\cite{dpmd2018} sGDML,\cite{Tkatch2018, Tkatch19} PhysNet,\cite{PhysNet} KREG,\cite{KREG} pKREG,\cite{pKREG} and KRR\cite{KREG}. We give brief descriptions of these methods below. As seen in that paper and also below, all the methods can reach RMS fitting errors for energies of 0.1 kcal~mol$^{-1}$ when trained just on energies.  However, in the time required for prediction (energies plus forces) there are differences of factors of ten or more. There are of course caveats to timings, but in this case, all timings were done on the same compute node, an Intel Xeon Gold 6240 2-Processor 24.75M Cache 2.60 GHz. A similar assessment of some of these ML methods was very recently made including the recent Atomic Cluster Expansion (ACE) method\cite{ACEPIP_21} using revised MD17 datasets.  

These methods have been described in detail in the two recent assessment papers \cite{dralchemsci, ACEPIP_21} and so we just give a brief description here.  In broad terms these methods  can be categorized into kernel-based approaches (e.g., GAP-SOAP, sGDML, KREG, pKREG, KRR) and neural network(NN)-based ones. The kernel-based approaches represent the potential energy as a linear combination of kernel functions that measure the similarity between the input molecular configuration and every configuration in the training set. As a result, the cost of prediction scales as $O(N)$, where $N$ is the size of the training data. The differences between these kernel based methods are the choice of the kernel function and the descriptor used to represent the molecular configuration. For example, in Kernel Ridge Regression (KRR), many descriptors, such as aligned Cartesian coordinates, Coulomb matrix \cite{cmatrix} (in KRR-CM), and RE descriptor \cite{KREG} (in KREG), have been used. Common kernel functions include the Gaussian kernel function. pKREG uses a  permutationally invariant kernel, and GAP-SOAP uses Smooth Overlap of Atomic Positions (SOAP) descriptor \cite{soap16} and kernel \cite{GP-2013}. sGDML slightly differs from the other methods as it directly learns the force by training in the gradient domain.\cite{Tkatch2018,Tkatch19}
The NN-based approaches studied in the paper by Pinbeiro et al. can be further divided into two families. ANI and DPMD can be viewed as variants of the Behler-Parrinello NN (BPNN) \cite{Behler15}: the potential energy is written as the sum of atomic ones, and the descriptor is a local one centered on each atom that describes the local environment of that atom. The descriptors used in ANI and DPMD are different, but they are both manually designed. On the other hand, the second family, ``message-passing'' NN, inspired by graph convolutional NN, learns the descriptor. The descriptor of an atom gets updated using information from its neighbors. PhysNet, SchNet,\cite{SchNet} and MEGNet\cite{2019Chen} are examples of this family. One advantage of NN-based methods over kernel-based ones is that the cost of prediction is independent of the size of training data once the NN structure is determined.
The ACE approach represents the potential as a body-ordered expansion (each atom is one ``body''), which is resummed into atomic energies. Each atomic energy is a linear combination of a set of permutationally-invariant basis functions centered on that atom.

In our opinion, computational efficiency is of primary importance for ML PESs to become transformative to the field. By transformative, we mean the leap from classical to quantum simulations of the dynamics and statistical mechanics of molecules, clusters and realistic samples for condensed phase systems. While classical simulations have been done for years using ``direct dynamics'' (slow compared to using a PES of course), this is simply not possible for quantum simulations.  For these one must have efficient ML PESs. For example, for a  diffusion Monte Carlo (DMC) calculation of the zero-point energy of an isolated molecule, roughly 10$^8$ or more energy evaluations can be required for convergence.\cite{tropolone20,glycine20,conte_glycine20,NandiQuBowman2019}  Nuclear quantum effects are known to be important for molecules, clusters, etc. with H atoms, and so incorporating these effects in simulations is important.

Here we add PIP to the list of ML methods mentioned above for the ethanol case study.  We use ethanol to showcase the performance of the new incorporation of reverse differentiation for gradients in our PIP software.  The details for this are given below, followed by a short demonstration for the 4-body water potential that we recently reported\cite{4body} and then the detailed analysis for ethanol. Finally, we present a new PES for ethanol that is based on a dataset of B3LYP energies and gradients that extend to much higher energies than the MD17 dataset.  The new PES is used in DMC calculations of the zero-point energy.  Such calculations fail using a precise fit to the MD17 dataset, which is severely limited by the sampling of energies from a 500 K MD simulation. 

\section*{Methods}
\subsection*{Recap of the Current PIP Approach }
\hspace{\parindent}In the PIP approach to fitting,\cite{Braams2009} the potential $V$ is represented in compact notation  by 
\begin{equation}
V(\bm{\tau})= \sum_{i=1}^{n_p} c_i p_i(\bm{\tau}),
\label{eq1}
\end{equation}
where $c_i$ are linear coefficients, $p_i$ are PIPs, $n_p$ is the total number of polynomials for a given maximum polynomial order, and $\bm{\tau}$ are transformed internuclear distances, usually either Morse variables, exp($-r_{n,m}/\lambda$),  or inverse distances, $1/r_{n,m}$, where $r_{n,m}$ is the internuclear distance between atoms $n$ and $m$. The range (hyper) parameter, $\lambda$, is usually chosen to be 2 bohr. The linear coefficients are obtained using standard least squares methods for a large data set of electronic energies at scattered geometries (and, for large molecules, using gradients as well). The PIPs, $p_i$ in Eq. (\ref{eq1}), are calculated from \emph{MSA} software\cite{Xie-nma, msachen} and depend on the monomials, $m_j$, which themselves are simple functions of the transformed internuclear distances $\bm{\tau}$. The inputs to the $MSA$ software include both the permutational symmetry and the overall order desired for the permutationally invariant polynomials. In brief the \emph{MSA} software proceeds by producing all monomials obtained from a seed monomial by performing the permutations of like atoms specified in the input.  Examples of this step are given in the review by Braams and Bowman.\cite{Braams2009} Then polynomials, which are formally the sum of monomials, are obtained in a recursive method starting with the lowest-order polynomials.  In this approach higher-order polynomials are obtained by a binary factorization in terms of lower order polynomials plus a remainder. Details are given elsewhere\cite{Xie-nma, msachen}; however, we mention this essential aspect of the software as it gives some insight into complexity of determining the gradient of this representation of the potential. 

For some applications, there are further requirements on the PIP basis set so that one can ensure that the fit rigorously separates into non-interacting fragments in asymptotic regions.  Identifying polynomials that do not have the correct limiting behavior is what we call ``purification''\cite{purified14, ConteQuBowman2015, purified15c, QuConteHoustonBowman2015} of the basis. Details of a recent example to the 4-body water PIP PES have been given; we refer the interested reader to that\cite{4body} and earlier references. Polynomials that do not have this property (``disconnected terms''\cite{purified13})
are labeled as  $q_i$ and separated from the basis set used to calculate the energy in Eq. (\ref{eq1}).  Thus, we now have polynomials of two types, those having the correct limiting behavior that will be used in the energy and gradient calculation, $p_i$ (see Eq. (\ref{eq1})), and those, $q_i$, that, while not having the correct limiting behavior, might still be needed because they could occur, for example, with multiplication by a polynomial that does have the correct limiting behavior. 

\subsection*{PIP Approach with Compaction and Fast Gradient Evaluation }
The first enhancement to the PIP approach is what we call ``compaction'' and is aimed at determining which polynomials $q_i$ and which monomials $m_i$ are not necessary.  We identify the unneeded monomials by increasing, one at a time, the value of the monomial to see if the values of the surviving $p_i$ polynomials change.  If they do not, that monomial may safely be eliminated.  We identify the unneeded $q_i$ by enumerating all needed $q_i$ that occur in the definitions of the $p_i$ as well as the $q_j$ with $j < i$ needed to define those $q_i$, and then taking the remainder to be unneeded.  The compaction then consists in deleting all references to the unneeded $m_i$ and $q_i$, followed by renumbering of all the surviving $m_i$, $q_i$, and $p_i$.  

The final steps in the development of the basis set are to add methods for calculating analytical gradients .  The first method, which we will call the ``normal'' gradient calculation,\cite{QuBowman2019,NandiQuBowman2019} is obtained formally by differentiating both sides of Eq. (\ref{eq1}) to determine how the change in potential depends on the change in the $p_i$.  Of course, the $p_i$ depend on $q_i$, $m_i$, and the internuclear distances, themselves a function of the atomic Cartesian coordinates.  Thus, we must also differentiate $q_i$, $m_i$, and $\tau_i$ with respect to each of the $3N$ Cartesian coordinates. We accomplish this conveniently by using the symbolic logic in Mathematica,\cite{Mathematica} a program whose mixture of text manipulation and expression evaluation is also used to write Fortran code for the aforementioned purification and compaction steps.  

Although the simple differentiation just described for the analytical gradients is straightforward, its implementation does not result in a fast gradient calculation. For example, the straightforward code would need to evaluate all the differentiated monomials, polynomials and $\bm{\tau}$ values $3N$ times for a single geometry.

We have also written a ``Fast (forward) Derivative'' method\cite{conte20_efficientPIP} that uses Mathematica's symbolic logic to solve for the derivatives of each $p_i$ with respect to each component of $\bm{\tau}$, which we denote by $\tau_M$, where  $M=1,N(N-1)/2$. We start with the derivatives of Eq. (\ref{eq1}):
In our case, we have 
\begin{equation}
\frac{\partial{V}}{\partial{\tau_M}} =  \frac{\partial{}}{\partial{\tau_M}}(\sum_{i=1}^{n_p}
c_i p_i) = \sum_{i=1}^{n_p}c_i \frac{\partial{p_i}}{\partial{\tau_M}}
\label{eq: dEdxi}
\end{equation}
Next let $\alpha_n$ be the $x$, $y$, or $z$ Cartesian coordinate of atom $n$. The calculation is completed by determining
\begin{equation}
\frac{\partial{V}}{\partial{\alpha_n}} =\frac{\partial{V}}{\partial{\tau_M}}\frac{\partial{\tau_M}}{\partial{\alpha_n}} =  \sum_{i=1}^{n_p}c_i \frac{\partial{p_i}}{\partial{\tau_M}}\frac{\partial{\tau_M}}{\partial{\alpha_n}}.
\label{eq: dEdxck}
\end{equation}
For any geometry, the derivatives $\frac{{\partial{p_i}}}{{\partial{\tau_M}}}$ on the rhs of Eq. (\ref{eq: dEdxck}) are stored so that the calculation of each $p_i$ derivative with respect to each $\bm{\tau}$ component does not need to be repeated; only the remaining $3N$ derivatives in Eq. (\ref{eq: dEdxck}) of the $\bm{\tau}$ components with respect to the Cartesian coordinates need to be performed.  In addition, since many of the derivatives are zero, we store only the non-zero ones and pair them with an index which indicates to which $p_i$ and $\tau_M$ they belong.  This method speeds up the calculation substantially but is still not the best method, which we describe next.

Automatic differentiation has been the subject of much current study\cite{BaydinPearlmutter2014,Baydin2018} and is widely disseminated on the internet. It has found its way into computational chemistry, \cite{autodiffSchaefer} mainly via libraries written in Python.

We have discussed above some of the issues involved in what is called the ``forward'' method of automatic differentiation.  When there are only a few inputs and many outputs, the forward method is usually quite adequate.  For our problem, however, there are many inputs ($3N$ Cartesian coordinates) and only one output (the energy, or its gradient).  In this case, a ``reverse'' (sometimes called ``backward'') differentiation is often faster.  In either case, we start with a computational graph of the steps to be taken in the forward direction that ensures that the needed prerequisites for any step are previously calculated and provides an efficient computational plan; i.e., does not recalculate anything that has been previously calculated.  The \emph{MSA} output provides such a plan for calculating the energy, which in our case is a fairly linear plan: to get the potential energy, start with the the Cartesian coordinates, $\alpha_n$, then calculate the transformed internuclear distances $\tau_M$, then the $m_k$, then the $p_i$ (or, in our purified case, the $q_j$ followed by the $p_i$), and finally take the dot product of the $c_i$ coefficients with the evaluated $p_i$ polynomials.  Note that, in principal, any of the quantities, $\alpha,\tau,m,q,$ or $p$, can influence any of the ones that go after it (in the forward direction).  In addition, note that there is a split at the end, so that, for example, any $p$ can influence the energy either by its contribution to the dot product or through its influence on any of the $p$ that come after it.  The sequence of steps in the correct order is, of course, maintained in the purification and compaction steps.  

For the forward derivatives, everything is the same as for the energy except that each step is replaced by its derivatives, as shown in the left column of Table \ref{tab: computationgraph}, which follows the Fortran notation of putting the subscripts in parentheses. We also note that Fortran code makes no distinction between full and partial derivatives.  The ``normal'' differentiation discussed in the previous paragraph is accomplished by working in the forward (up) direction of the left column, but one has to make $3N$ forward passes of the computational plan to get the gradients.  The reverse automatic differentiation allows one to work backwards (down in the right column) from the derivative of the potential energy to get all $3N$ gradients in one pass of the computational plan, after having run the energy plan once in the forward direction.  The strategy results in an extremely efficient method. It also scales more favorably with an increase in the number of atoms because, in the reverse direction, the cost of the $3N$ gradients is typically 3--4 times the cost of the energy,\cite{Baydin2018,GriewankWalther2008} whereas in the forward direction it is about $3N$ times the cost of the energy.  Evidence that this is the case for application to PIPs will be presented in a subsequent section.

\begin{table}[ht]

\caption{Forward and Reverse Automatic Differentiation for PIP basis sets}
\centering
\begin{tabular*}{\columnwidth}{ c c c }
\hline
\hline\noalign{\smallskip}

Forward (up) &\hspace{.5cm} &Reverse (down) \\
${\partial{V}} = \bm{c}\cdot\partial{\bm{p}}$ & & ${\partial{V}} = \bm{c}\cdot\partial{\bm{p}}$\\
\noalign{\smallskip}\hline\noalign{\smallskip}
$dp(n_p) = $... & & $a(n_{max})= \frac{\partial{V}}{\partial{p(n_p})} = c(n_p) $\\
$dp(n_p-1) = $... & & $a(n_{max}-1)=$ \\
 & &$c(n_p-1) + a(n_{max})\frac{\partial{p(n_p)}}{\partial{p(n_p-1)}}  $\\
... & & ...\\
$dp(0) = dm(0)$& & \\
$dq(n_q) = $ ... &  &\\
... & &...\\
$dq(1) = $ ... & &\\
$dm(n_m) = $ ...& & \\
... & & ... \\
$dm(0) = 0$ & & $a(j)=\sum_{i=j+1}^{n_{max}}a(i)\frac{\partial{t(i)}}{\partial{m(0)}}$\\
... & & ...  \\
$d\tau(M) =$ ... & & \\
... & & ...  \\
$dx_n= dx_1 = $... & & $a(3N) = \frac{\partial{V}}{\partial{x_1}}$\\
$dy_n= dy_1  = $... & & $a(3N-1) = \frac{\partial{V}}{\partial{y_1}}$\\
... & & ...  \\
$dz_n= dz_N = $... & & $a(1) = \frac{\partial{V}}{\partial{z_N}}$\\
\noalign{\smallskip}\hline
\hline\noalign{\smallskip}
\end{tabular*}
\label{tab: computationgraph}
\end{table}

We define the adjoint, $a_j$,  of a particular instruction as the partial derivative of the potential energy with respect to the conjugate variable, $t_j$, where $dt_j$ is the differential that is defined by the instruction in the forward direction: thus, $a_j = \frac{\partial{V}}{\partial{t_j}}$. The adjoints will provide the instructions for proceeding in the reverse direction, down column two of Table \ref{tab: computationgraph}.  When we reach the end, the adjoints $\frac{\partial{V}}{\partial{\alpha_{n}}}$ will give the $3N$ derivatives we seek. Of course, in determining which $t_j$ contribute to the adjoint, we must sum all the ways that a change in $V$ might depend on $t_j$.

It is instructive to work a few adjoints in the reverse direction (see Table \ref{tab: computationgraph}). The first equation moving in the reverse direction will be the adjoint of the conjugate variable $dp_{n_p}$, defined the forward direction, so the adjoint to evaluate is $\frac{\partial{V}}{\partial{p_{n_p}}}$ (which is equal to $c_{n_p}$).

The next line in the reverse direction defined $dp_{(n_p-1)}$ in the forward direction. The change in $V$ with $p_{n_p-1}$ now depends potentially both on how a change in $p_{n_p-1}$ influences $V$ indirectly through $p_{n_p}$ and on how it influences $V$ directly through the contribution to the dot product.  Thus, the adjoint is  $\frac{\partial{V}}{\partial{p_{n_p-1}}}=c_{n_p-1} +\frac{\partial{V}}{\partial{p_{n_p}}}\frac{\partial{p_n}}{\partial{p_{n_p-1}}}=c_{n_p-1} +a(n_{max})\frac{\partial{p_n}}{\partial{p_{n_p-1}}}$.

For the third line (not shown in the table), the adjoint is $\frac{\partial{V}}{\partial{p_{(n_p-2)}}}$. The change in $V$ with $p_{n_p-2}$ depends potentially both on how a change in $p_{n_p-2}$ influences $V$ indirectly through $p_{n_p}$ and $p_{n_p-1}$ and on how it influences $V$ directly through the contribution to the dot product. Thus, the adjoint is $\frac{\partial{V}}{\partial{p_{n_p-2}}}=c_{n_p-2}+ \frac{\partial{V}}{\partial{p_{n_p-1}}} \frac{\partial{p_{n_p-1}}}{\partial{p_{n_p-2}}} +\frac{\partial{V}}{\partial{p_{n_p}}}\frac{\partial{p_{n_p}}}{\partial{p_{n_p-2}}}=c_{n_p-2}+
a(n_{max}-1) \frac{\partial{p_{n_p-1}}}{\partial{p_{n_p-2}}} + \\a(n_{max})\frac{\partial{p_{n_p}}}{\partial{p_{n_p-2}}}$.

Notice in all cases that the adjoint we seek is the $c$ coefficient of the conjugate variable plus the sum of all adjoints that preceded it, each times the derivative of its conjugate variable with respect to the conjugate variable of the adjoint we seek.  The direct contribution through the $c_i$ occur only if the conjugate variable is a $p$.  This observation gives the formula for assigning all the remaining adjoints: 
\begin{equation}
a_j(t_j)= c_i \delta_{t,p} +\sum_{i=j+1}^{i_{max}} a_i \frac{\partial{t_i}}{\partial{t_j}},
\label{eq: adjointrule}
\end{equation}
where $a_j$, with conjugate variable $t_j$, is the adjoint to be calculated, $i_{max}$ is the maximum number of adjoints, $c_i$ is the coefficient associated with $p_i$ when $t_j=p_i$,  and $\delta_{t,p}$ is 1 if $t$ is a $p$ and 0 otherwise.  Two ``toy'' examples, one of a homonuclear diatomic molecule and one of a single water molecule, are worked in detail in the Supplementary Material.

In the Mathematica implementation of our software, we pursue two routes in parallel: the first is to evaluate symbolically the adjoints in Eq. (\ref{eq: adjointrule}) using Mathematica code, and the second is to create the Fortran code from the Mathematica code for the same adjoints.  The symbolic logic of Mathematica is used to solve the partial derivative factor of the adjoint terms.  The resulting adjoint is turned into Fortran format and saved as an instruction list.  
Because many of the partial derivatives on the rhs of Eq. (\ref{eq: adjointrule}) are zero, it is fastest to enumerate all the $t_i$ with $i>j+1$ that depend on $t_j$ and then perform the terms in the sum only for those values of $i$. The key Mathematical program for evaluating a particular adjoint is provided in the Supplementary Material, as is a brief description of the various Mathematica steps involved in fragmentation, purification, compaction, adding or pruning polynomials, and appending derivative functions.

We need to make one point clear: the Mathematica code must be run to generate Fortran output for each permutational symmetry in which the user might be interested.  Thus, there is an overhead on the order of an hour or so to generate the fast derivative method for most problems of interest.  Once the basis set and derivative method have been established, however, they can be run without further change. Also, the basis and reverse derivative code can be used for any molecule with the same permutational symmetry.  

\section*{Results}
\subsection*{4-body water interaction potential}
The first set of results is for the 12-atom, 4-body water potential, which we recently reported.\cite{4body} Here, we used 
permutational symmetry 22221111 with a total polynomial order of 3.  The \emph{MSA} basis was purified by distancing each of the four monomers, one at a time, and distancing each of the sets of dimers, two at a time. In Table \ref{tab: timing} we show the times in seconds for the calculation of the energy and the $3N = 36$ gradients for purified/non-compacted and purified/compacted basis sets using four different gradient methods. Each time is for the evaluation of 20,000 geometrical configurations.  It is clear from the table that the reverse derivative method is fastest and that it runs about 17 times faster than the 2-point finite difference method, often used in molecular dynamics calculations.  In addition, compaction of the purified basis gives a further speed-up in this case of about 40\%. (Future plans call for using the 4-b PES for a full \emph{ab initio} water potential, so having a fast gradient is important for the usual MD and possibly PIMD simulations.)

\begin{table}[ht]
\caption{Total time for performing 20 000 gradient sets ($3N = 36$ gradients each) for a 22221111 permutational symmetry basis of maximum order 3 and various derivative methods. This is a 12-atom basis, which was used recently for the 4-b water potential.\cite{4body}}

\begin{tabular*}{\columnwidth}{ l c c c c }
\hline
\hline\noalign{\smallskip}

  & 2-pt. Finite & Normal & Fast & Reverse \\
   & Difference & anaytical & Derivative & Derivative \\
\noalign{\smallskip}\hline\noalign{\smallskip}
Fully purified/ & 13.2 s & 9.7 s & 1.0 s & 0.7 s \\
non-compact & & & &  \\
 & & & & \\
Fully purified/ & 3.2 s & 2.0 s & 0.8 s & 0.2 s \\
compact & & & & \\
\noalign{\smallskip}\hline
\hline\noalign{\smallskip}
\end{tabular*}
\label{tab: timing}
\end{table}

\subsection*{Ethanol }
\subsubsection*{Assessment of ML Methods}
As noted already, the performance of a number of ML methods was examined in detail for ethanol, using the MD17 database of energies and forces.\cite{MD17} This assessment provides detailed results with which we can compare our PIP method.  The comparison is done using the same protocol used previously,\cite{dralchemsci}  namely to obtain the RMSE for energies and gradients using training data sets of increasing size and also for training on just energies or on energies plus gradients. The permutational symmetry we use here is 321111, and we also consider the performance of two PIP bases, one order of 3 and the other of order 4.  These have 1898 and 14 752 terms, respectively.  We also consider a third basis of size 8895, obtained from pruning the n = 4 one. The procedure to do this is straightforward and is the following.  The desired number of terms is the input, and all the polynomials for n = 4 (14 752) are evaluated using the maximum values of the Morse variables (taken over the data set). Then the desired number of polynomials is obtained by starting with the one of largest value and proceeding downward.  This procedure can reduce a basis to any desired size. 

\begin{figure}[htbp!]
    \centering
    \includegraphics[width=\columnwidth]{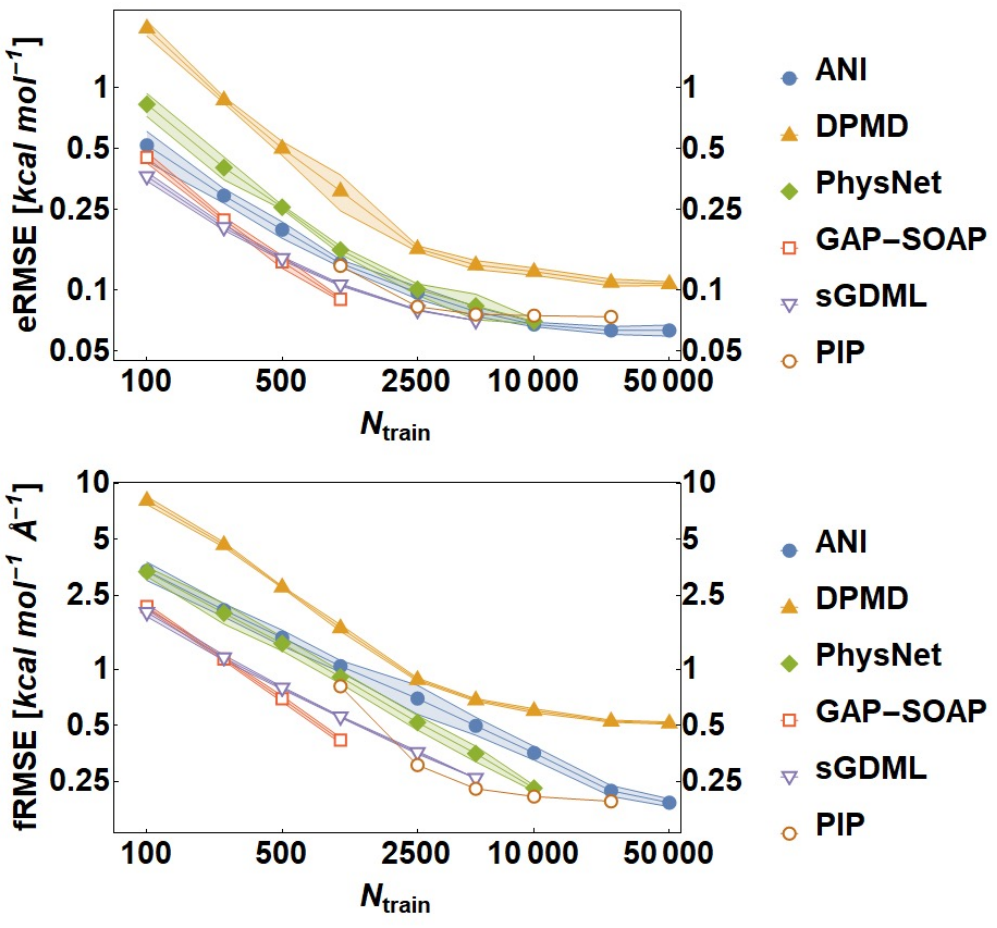}
    \caption{Comparison of the different machine learning potentials trained on energies and forces for the MD17-ethanol dataset. The PIP results are for a basis with 14752 terms. The upper panel shows root mean-
squared error in energies (eRMSE) vs the number of training points and the lower panel shows root mean-
squared error in forces (fRMSE) vs the number of training points.}
    \label{fig:Ethanol_fitting}
\end{figure}

\begin{figure}[htbp!]
    \includegraphics[width=\columnwidth]{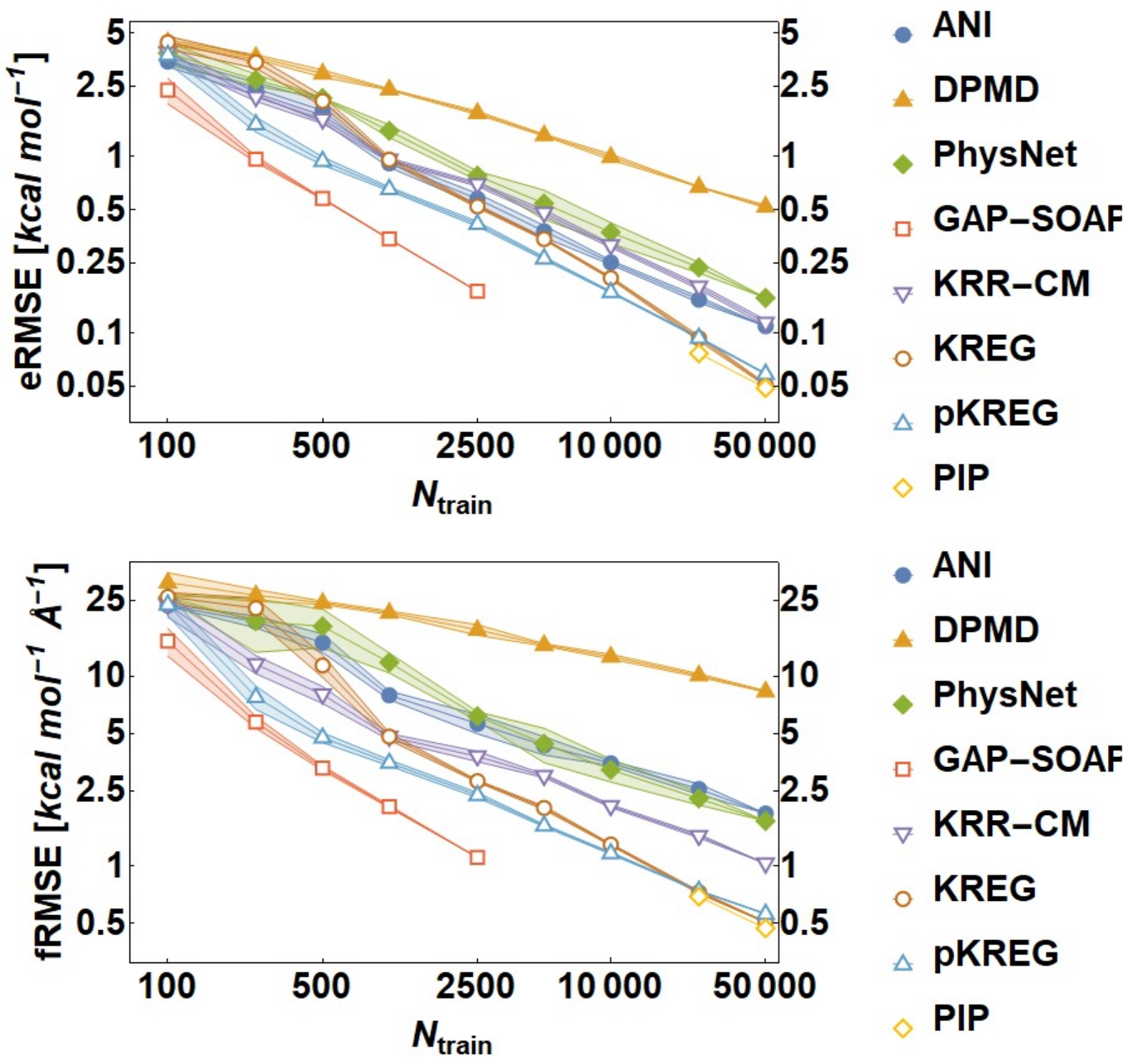}
    \caption{Comparison of the different machine learning potentials trained on energies only for the MD17-ethanol dataset. The PIP results are for a basis with 14752 terms. The upper panel shows root mean-squared error in energies (eRMSE) vs the number of training points and the lower panel shows root mean-squared error in forces (fRMSE) vs the number of training points.}
     \label{fig:Ethanol_fitting2}
\end{figure}

\begin{figure}[htbp!]
    \includegraphics[width=\columnwidth]{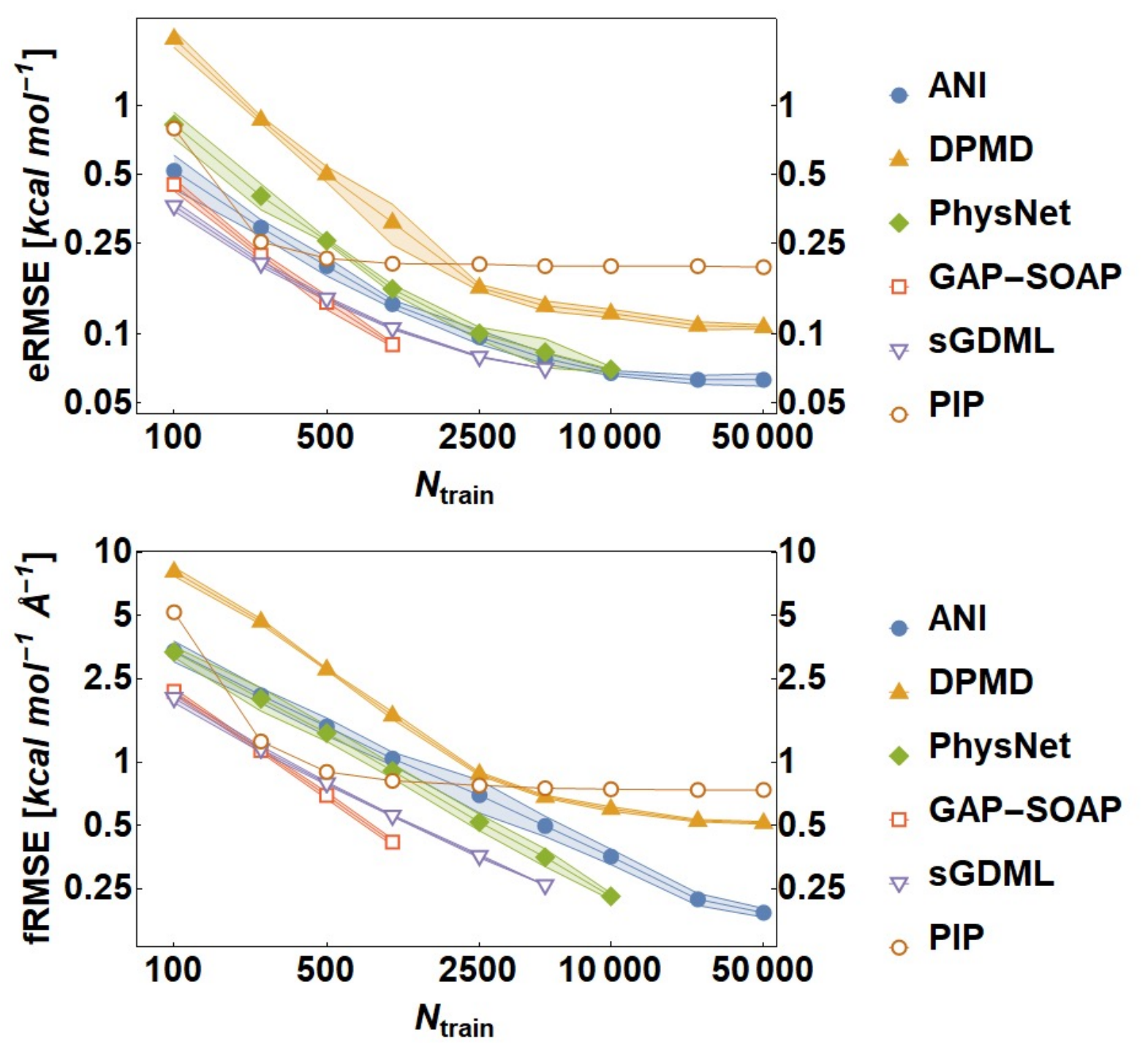}
    \caption{Comparison of the different machine learning potentials trained on energies and forces for the MD17-ethanol dataset. The PIP results are for a basis with 1898 terms. The upper panel shows root mean-squared error in energies (eRMSE) vs the number of training points and the lower panel shows root mean-squared error in forces (fRMSE) vs the number of training points.}
        \label{fig:Ethanol_fitting3}
\end{figure}

\begin{figure}[htbp!]
    \includegraphics[width=\columnwidth]{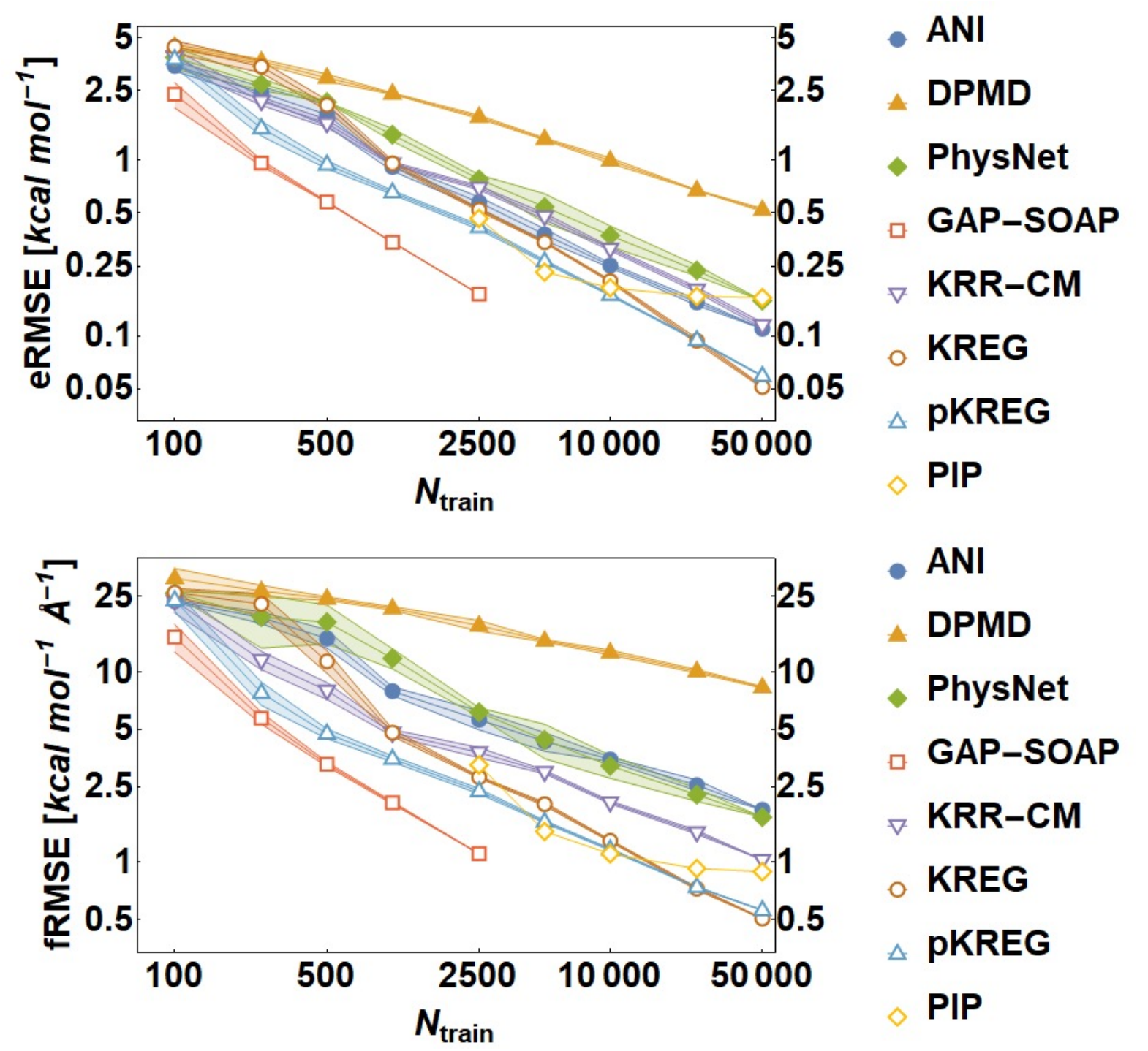}
    \caption{Comparison of the different machine learning potentials trained on energies only for the MD17-ethanol dataset. The PIP results are for a basis with 1898 terms. The upper panel shows root mean-squared error in energies (eRMSE) vs the number of training points and the lower panel shows root mean-squared error in forces (fRMSE) vs the number of training points.}
        \label{fig:Ethanol_fitting4}
\end{figure}

\begin{table}[htbp!]
\caption{Comparison of prediction time of the different machine learning potentials trained on MD-17 ethanol energies and forces. The times listed are those for calculation of the energy and forces for a total of 20 000 geometric configurations.}
\label{tab:train1}

\begin{threeparttable}
	\begin{tabular*}{\columnwidth}{@{\extracolsep{\fill}}rccccccc}
	\hline
	\hline\noalign{\smallskip}
	  & \multicolumn{7}{c}{Prediction Timing (sec)} \\
    \noalign{\smallskip} \cline{2-8} \noalign{\smallskip}
     $N_{\rm train}$ & ANI & DPMD & Phys & GAP- & sGDML & PIP\tnote{a} & PIP\tnote{b} \\
      & & & -Net & SOAP & & &  \\
	\noalign{\smallskip}\hline\noalign{\smallskip}
     100   & 27 & 180 & 213 & 212  & 7   &  0.23 & --- \\
     250   & 27 & 176 & 215 & 307  & 10  &  0.23 & --- \\
     500   & 27 & 176 & 214 & 560  & 15  &  0.23 & --- \\
     1000  & 27 & 182 & 215 & 1100 & 25  &  0.23 & 2.3 \\
     2500  & 27 & 189 & 216 & --- & 63  &  0.23 & 2.3 \\
     5000  & 27 & 186 & 216 & --- & 195 &  0.23 & 2.3 \\
     10000 & 27 & 188 & 213 & --- & --- &  0.23 & 2.3 \\
     25000 & 27 & 179 & 215 & --- & --- &  0.23 & 2.3 \\
     50000 & 27 & 176 & 214 & --- & --- &  0.23 & 2.3 \\
	\noalign{\smallskip}\hline
	\hline
   
	\end{tabular*}
   
   \begin{tablenotes}
   \item[a] Maximum polynomial order 3 is used to fit the data, which leads to 1898 PIP bases.
   \item[b] Maximum polynomial order 4 is used to fit the data, which leads to 14572 PIP bases.
   \end{tablenotes}

\end{threeparttable}
\end{table}

\begin{table}[htbp!]
\caption{Comparison of prediction time of the different machine learning potentials trained on MD-17 ethanol energies only. The times listed are those for calculation of the energy and forces for a total of 20 000 geometric configurations.}
\label{tab:train2}

\begin{threeparttable}
	\begin{tabular*}{\columnwidth}{@{\extracolsep{\fill}}rcccccccc}
	\hline
	\hline\noalign{\smallskip}
	  & \multicolumn{8}{c}{Prediction Timing (sec)} \\
    \noalign{\smallskip} \cline{2-9} \noalign{\smallskip}
     $N_{\rm train}$ & ANI & DPMD & Phys & GAP & KRR & pKREG & PIP\tnote{a} & PIP\tnote{b} \\
      & & & -Net & SOAP & -CM & &  &   \\
	\noalign{\smallskip}\hline\noalign{\smallskip}
     100   & 26 & 99  & 313 & 211  & 2    & 2  & --- & --- \\
     250   & 27 & 95  & 291 & 306  & 5    & 3  & --- & --- \\
     500   & 27 & 95  & 288 & 561  & 11   & 5  & --- & --- \\
     1000  & 26 & 101 & 304 & 1101 & 21   & 10  & --- & --- \\
     2500  & 26 & 94  & 294 & 3611 & 52   & 22  & 0.23 & --- \\
     5000  & 26 & 97  & 304 & --- & 102  & 49  & 0.23 & --- \\
     10000 & 26 & 97  & 301 & --- & 203  & 97  & 0.23 & --- \\
     25000 & 26 & 99  & 298 & --- & 508  & 227 & 0.23 & 2.3 \\
     50000 & 26 & 93  & 295 & --- & 1018 & 438 & 0.23 & 2.3 \\
	\noalign{\smallskip}\hline
	\hline
   
	\end{tabular*}
   
   \begin{tablenotes}
   \item[a] Maximum polynomial order 3 is used to fit the data, which leads to 1898 PIP bases.
   \item[b] Maximum polynomial order 4 is used to fit the data, which leads to 14572 PIP bases.
   \end{tablenotes}
  
\end{threeparttable}
\end{table}

\begin{table}[htbp!]
\centering
\caption{Assessment of the different machine learning potentials trained on energies for the MD17-ethanol dataset. The upper row shows root mean-squared error energy targets of around 0.5 and 0.1 kcal~mol$^{-1}$ for a test set of 20 000 configurations, while the columns show, for each of the methods, the training size required to achieve the targets and the time required for 20 000 energy and forces predictions. Timings based on the same Intel Xeon Gold 6420 processor, see text for details.}
\label{tab:compare}

\begin{threeparttable}
	\begin{tabular*}{\columnwidth}{@{\extracolsep{\fill}}lrrrr}
	\hline
	\noalign{\smallskip}
	\hline

	\hline
	\noalign{\smallskip}
     Target eRMSE & \hspace{.7 cm}0.5\hspace{.1 cm} & \hspace{.7 cm}0.1\hspace{.1 cm} & \hspace{.7 cm}0.5\hspace{.1 cm} & \hspace{.7 cm}0.1\hspace{.1 cm} \\
 (kcal-mol$^{-1}$) & & & & \\
 \hline
 	\noalign{\smallskip}
    \textbf{Method:} & \multicolumn{2}{c}{\textbf{\hspace{.4 cm}Training }} & \multicolumn{2}{c}{\textbf{\hspace{.4 cm}Prediction}} \\
   & \multicolumn{2}{c}{\textbf{\hspace{.4 cm}Size}} & \multicolumn{2}{c}{\hspace{.4 cm}\textbf{Time\tnote{a} (sec)}} \\
 \hline
 	\noalign{\smallskip}
  pKREG & 2500 & 25000 & 22 & 227 \\ 
 KRR & 5000 & 50000 & 102 & 508 \\ 
 sGDML\tnote{b} & 100 & 1000 & 7 & 25 \\ 
 GAP-SOAP & 500 & 2500 & 561 & 3611 \\ 
 ANI & 2500 & 50000 & 26 & 26 \\ 
 PhysNet & 5000 & 50000 & 300 & 300 \\ 
 PIP\tnote{c} & 2500 & 10000 & 0.23 & 0.23 \\ 
 PIP\tnote{d} & N/A & 25000 & N/A & 2.3 \\ 
	\hline
   	\noalign{\smallskip}\hline
	\end{tabular*}
   
   \begin{tablenotes}
   \item[a] Energies and forces (20 000 configurations).
   \item[b] Trained on forces only.
   \item[c] 1898-term basis.
   \item[d] 14 572-term basis. 25 000 is the smallest training size (see text for details).
   \end{tablenotes}

\end{threeparttable}
\end{table}

Fig. \ref{fig:Ethanol_fitting} shows a comparison of the root-mean-square (RMSE) values for the energies (eRMSE) and  forces (fRMSE) for the indicated methods as a function of the size of the training set, based on fits to energies and gradients, with the exception of the sGDML method, which was fit to gradients only. For this PIP fit, the basis set contains 14 752 terms.  All methods achieve small RMSE values at sufficiently high training sizes; the GAP-SOAP, sGDML and PIP methods converge more rapidly.  Similar plots for fitting to energies only are shown in Fig. \ref{fig:Ethanol_fitting2}.  Where the results are available (at high training numbers), the PIP and pKREG precision for both energies and forces are the best.  Note that with energies only, because of the large number of coefficients, it is inadvisable to fit the large-basis PIP set to train on data sets that have fewer than about 25 000 configurations because of likely overfitting.

Comparable figures to Figs. \ref{fig:Ethanol_fitting} and \ref{fig:Ethanol_fitting2} are shown for the smaller basis set (1898 terms) in Figs. \ref{fig:Ethanol_fitting3} and \ref{fig:Ethanol_fitting4}. 
As seen, training on energies plus gradients yields essentially the ultimate eRMSE and fRMSE for a training size of 1000 configurations. Although the precision is not as high as for the larger PIP basis and for other ML methods the timing results are much faster, especially for the non-PIP methods, as will be presented next. Training just on energies with this PIP basis does result in a smaller ultimate eRMSE. The results for using the PIP basis of size 8895, obtained from pruning the $n = 4$ one, to fit a dataset of 10 000 configurations are as follows. Training is done on energies plus gradients and produces eRMSE and fRMSE for this training dataset of 0.09 kcal~mol$^{-1}$ and 0.30 kcal~mol$^{-1}$~\AA$^{-1}$, respectively. The testing at 20 000 geometries produces eRMSE and fRMSE of 0.09 kcal~mol$^{-1}$ and 0.34 kcal~mol$^{-1}$~\AA$^{-1}$, respectively. Thus, fitting accuracy of this pruned basis is very similar to the large PIP basis.

We now consider the time required for calculation of the energies and gradients as a function of training size. An analysis of timing was also reported in a plot in ref. \citenum{dralchemsci} without the current PIP results. Tables \ref{tab:train1} and \ref{tab:train2} present timing results for different machine learning potentials trained on ethanol energies plus gradients or energies alone, respectively. For all conditions, the time required for calculation of 20 000 geometric configurations is far less than that for other methods listed, in most cases by more than one order of magnitude, particularly at the higher RMSE accuracy provided by larger training sizes. A note on the timing results is in order.  All of the results in Tables \ref{tab:train1} and \ref{tab:train2} were obtained with the same Intel processor (Xeon Gold 6240).   
A summary comparison of these results is provided in Table \ref{tab:compare}.  This shows the training size necessary to achieve the eRMSE target values of around 0.5 and 0.1 kcal~mol$^{-1}$ as well as the calculation time required for 20 000 energies and gradients. In order to reach a value of around 0.1 kcal~mol$^{-1}$ eRMSE, although a larger training size is necessary, the time required is approximately 50 times less for the small PIP basis and 5 times less for the large one, as compared to the fastest alternative.  We note that, while the training size is important, once one has the method in place, what matters to most users is how fast one can perform molecular dynamics and quantum calculations using the PES.

Very recently, the ACE method has been used to fit the ethanol MD17 data set, as well as datasets for other molecules.\cite{ACEPIP_21}  This method was trained and tested on splits of 1000 configurations each (energies plus gradients).  The ACE method achieves an MAE from around 0.1 kcal~mol$^{-1}$ to a low value of 0.03 kcal~mol$^{-1}$, depending on the values of the hyperparameters in this method.  A detailed comparison with the small and large basis PIP fits is given in Table \ref{tab:MAE}.  The timings for ACE were obtained on a 2.3 GHz Intel Xeon Gold 5218 CPU, which has essentially the same single-core performance as the Intel Xeon Gold 6240, but slower multi-core performance due to smaller number of cores (16 vs 18). Taking that into account we find that for comparable MAEs the PIP PESs run at factors between roughly 20 and 100 times faster than the ACE fits.

\begin{table}[htbp!]
\caption{Mean absolute errors (MAE) of energies (kcal-mol$^{-1}$) and forces (kcal-mol$^{-1}$\AA$^{-1}$) for ACE and PIP fits to MD17 datasets of energies and forces for ethanol, along with corresponding timings for 20 000 evaluations of energies and forces. Timings based on two Intel processors that run at about the same speed, see text for details.}
\label{tab:MAE}

\begin{threeparttable}
	\begin{tabular*}{\columnwidth}{@{\extracolsep{\fill}}rccc}
	\hline
	\hline\noalign{\smallskip}
     Method & eMAE & fMAE & Timing (sec) \\
	\noalign{\smallskip}\hline\noalign{\smallskip}
     ACE          & 0.10 & 0.40 & 29 \\
     PIP\tnote{a} & 0.15 & 0.50 & 0.23 \\
     ACE          & 0.06 & 0.30 & 65 \\
     PIP\tnote{b} & 0.06 & 0.12 & 2.3 \\
	\noalign{\smallskip}\hline
	\hline
   
	\end{tabular*}
   
   \begin{tablenotes}
   \item[a] 1898-term basis.
   \item[b] 14 572-term basis.
   \end{tablenotes}
  
\end{threeparttable}
\end{table}

\subsubsection*{A New ``DMC-certified'' PES}

The MD17 dataset for ethanol has been used to compare the performance of the ML methods with respect to training and testing RMS errors and prediction timings.  This dataset has been used for this purpose for a number of molecules.\cite{dralchemsci,Tkatchjcp,ACEPIP_21} Beyond this important utility, one can inquire about the many uses that the PES fits can be put to.  

In the case of ethanol one important application would be to get a rigorous calculation of the partition function.  This is complicated owing to the  coupled torsional modes, as pointed out in an approximate state-of-the-art study that, without a PES, relied on a variety of approximations to obtain the partition function.\cite{ethpartit} Another application, already noted above, is a diffusion Monte Carlo calculation of the zero-point energy and wavefunction.  For such calculations the dataset must have the wide coverage of the configuration space and corresponding energies. 
As we show below the MD17 ethanol dataset is distributed over the energy range from 0--12000 cm$^{-1}$ with respect to the minimum energy. This energy range is not sufficient to describe the zero-point energy, which is estimated to be roughly 17,500 cm$^{-1}$ from a normal mode analysis.  To emphasize this, we used the large basis PIP PES in DMC calculations. As expected, we encounter a large number of ``holes'', i.e., configurations with large negative energy relative to the minimum in the data base. These holes occurred mainly at regions of high energy,  where the MD17 dataset does not have coverage.

\begin{figure}[htbp!]
    \includegraphics[width=\columnwidth]{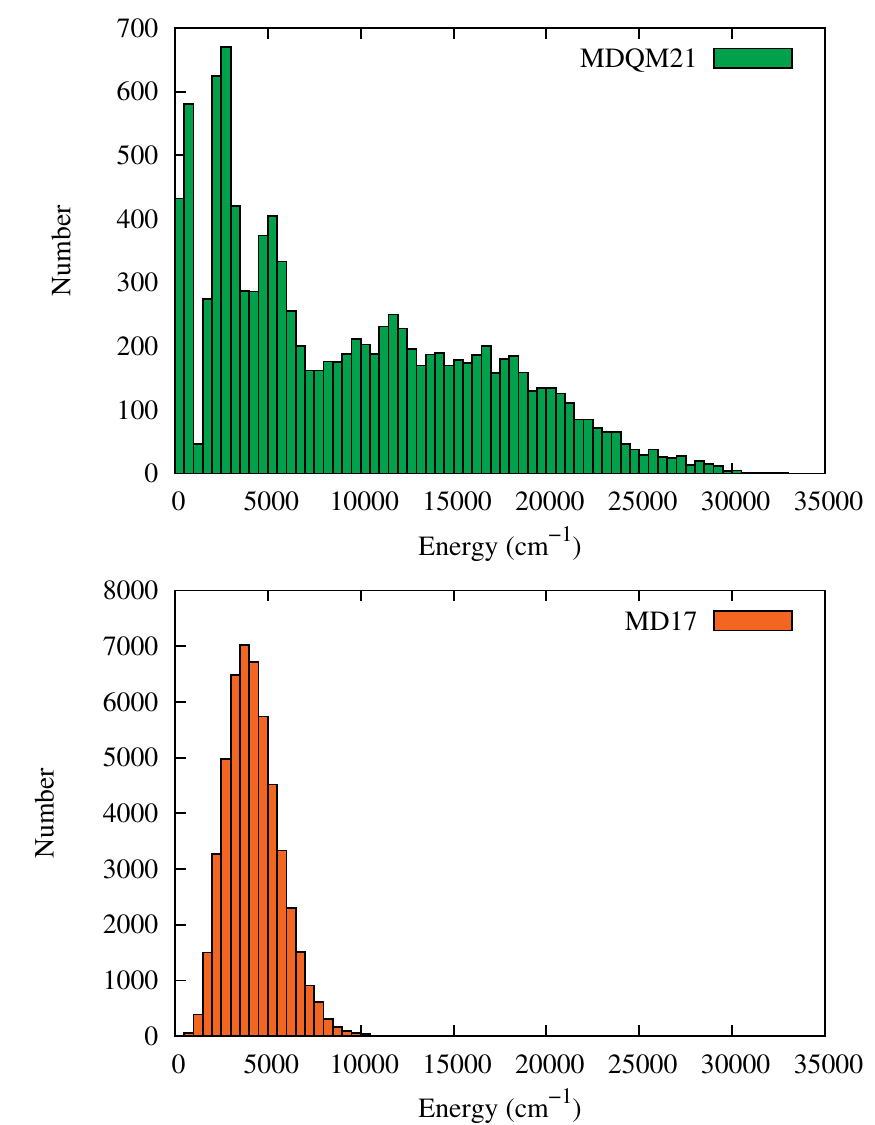}
    \caption{Distributions of electronic energies (cm$^{-1}$) of MDQM21 and MD17 dataset relative to their minimum value.}
        \label{fig:dataset_comparison}
\end{figure}

To address this problem, we generated a new dataset at the B3LYP/6-311+G(d,p) level of theory that has much larger coverage of configuration space and energies. The data sets of energies and gradients were generated using our usual protocol, namely \textit{ab initio} microcanonical molecular dynamics (AIMD) simulations at a number of total energies. These AIMD trajectories were propagated for 20000 time steps of 5.0 a.u. (about 0.12 fs) and with total energies of 1000, 5000, 10 000, 20 000, 30 000, and 40 000 cm$^{-1}$. A total of 11 such AIMD trajectories were calculated; one trajectory at the total energy of 1000 cm$^{-1}$ and two trajectories for each remaining total energies. The geometries and their corresponding 27 force components were recorded every 20 time steps from each trajectory to generate this new fitting dataset. These calculations are done using the Molpro quantum chemistry package.\cite{MOLPRO_brief} The final data set consists of 11 000 energies and corresponding 297 000 forces. We denote this new dataset as ``MDQM21". The distributions of electronic energies of this MDQM21 and MD17 datasets are shown in Fig. \ref{fig:dataset_comparison}. 

As seen there, the distribution of the MD17 dataset is the one that can be anticipated for 3$N$-6 classical harmonic oscillators at a thermal distribution at 500 K. For ethanol there are 21 such oscillators and the average potential is roughly 10 kcal~mol$^{-1}$ (roughly 3500 cm$^{-1}$), in accord with the distribution seen. By contrast, the MDQM21 dataset is very broad compared to that of the MD17 dataset. This is a direct result of running sets of direct-dynamics trajectories.

For the PES fits we divided the MDQM21 dataset into a training set of 8500 geometries and a test dataset of 2500 geometries. We used the same large PIP basis to fit this dataset using 80 percent for training and 20 percent for testing. The training RMSEs for energies and forces are 40 cm$^{-1}$ (0.114 kcal~mol$^{-1}$) and 0.000334 hartree bohr$^{-1}$ (0.396 kcal~mol$^{-1}$~{\AA}$^{-1}$), respectively. The testing RMSE for energies and forces are 51 cm$^{-1}$ (0.145 kcal~mol$^{-1}$) and 0.000484 hartree~bohr$^{-1}$ (0.574 kcal~mol$^{-1}$~{\AA}$^{-1}$), respectively. These are very similar to energy and force RMSEs for the large-basis PIP PES trained on the MSD17 dataset.

The new PES was  used successfully in DMC calculations. Each DMC trajectory was propagated for 25 000 time steps using 10 000 random walkers with the step size of 5.0 au. Fifteen DMC simulations were done, and the final ZPE, 17168 cm$^{-1}$, is the average of the 15 trajectories with the standard deviation of 12 cm$^{-1}$. The DMC calculations completed without any holes and the PES ``earns'' the title ``DMC-certified''. 

We also applied this PES for geometry optimization and normal-mode analysis and the agreement with direct calculations is excellent. Results are given in Supplementary Material.

\section*{Discussion}
\subsection*{ML Assessments}
We have shown for  ethanol, the PIP timings are 10 to 1000 times faster than most other widely-cited ML methods considered in a previous comprehensive assessment.\cite{dralchemsci}  Similarly large factors were reported earlier in comparison of timings with a straightforward GPR approach just fitting energies but using low-order PIPs as inputs and using Morse variables for energies of four molecules, including 10-atom formic acid dimer.  At roughly the same RMS error for energies (0.1 kcal~mol$^{-1}$ or less) the GPR method was factors of 10--50 or more slower than PIP run on the same compute node.\cite{PIP-GP}   A second example is 15-atom acetylacetone (AcAc). We recently reported timing for energies on  a 4-fragment PIP PES  of 0.08 ms per energy,\cite{QuAcAc} using a dataset of MP2 energies and gradients reported earlier to obtain a PhysNet PES for AcAc.\cite{meuwly2020} Timings were not reported on the PhysNet PES for AcAc; however, the time per energy was reported as 4 ms for the smaller molecule malonaldyde (Intel Core i7-2600K).\cite{meuwly2020} This is a factor of 50 larger than for the PIP PES and so consistent with the factor of 100 for larger basis PIP and 1000 for small basis seen for ethanol for PhysNet. A final example is 5-atom \ce{OCHCO+}, where a PIP-PES\cite{OCHCO15} is roughly 1000 times faster than a PES obtained with SchNet\cite{SchNet} and using the PIP-PES CCSD(T) electronic energies and run on the same Intel compute node. That ML method was tested on small molecules where PIP-PESs were previously reported.\cite{brorsen19}  

Thus, we conclude from a variety of tests, especially those presented here, that PIP PESs are significantly more computationally efficient for energies and now also for gradients than other ML methods, and we can ask why this might be so.  The short answer is the simplicity of Eq. \ref{eq1}. This is just a dot product of the expansion coefficients times low-order polynomials. These are obtained using a bi-factorization approach that is also efficient.\cite{Xie10,QuConteHoustonBowman2015} The training time using this approach is also quite efficient since it relies on solving the standard linear least-squares system of equations. A caveat about overall efficiency is the  additional overhead, relative to other ML methods, due to the time needed to generate the PIPs using the \emph{MSA} software. In the present case the time requires for the small PIP basis is a few minutes and for the large basis roughly 1 hr.  Clearly, these bases could be stored in a library of PIP bases for the given 9-atom symmetry and used for any other molecule with this symmetry.  However, given the small computational effort to get these basis, it's not clear that this is needed.

Finally, we note that the codes developed for the methods tested previously\cite{dralchemsci} and here use a variety of languages, e.g., Python, FORTRAN, C, Julia.  These were presumably selected by developers of the codes based on their specific criteria.  For scientific uses, especially for quantum calculations, which is the emphasis here,  computational speed is a high priority.  As already noted in the Introduction and seen here,  this is one aspect that clearly separates the performance of the codes.    

\begin{figure}[htbp!]
    \centering
    \includegraphics[width=\columnwidth]{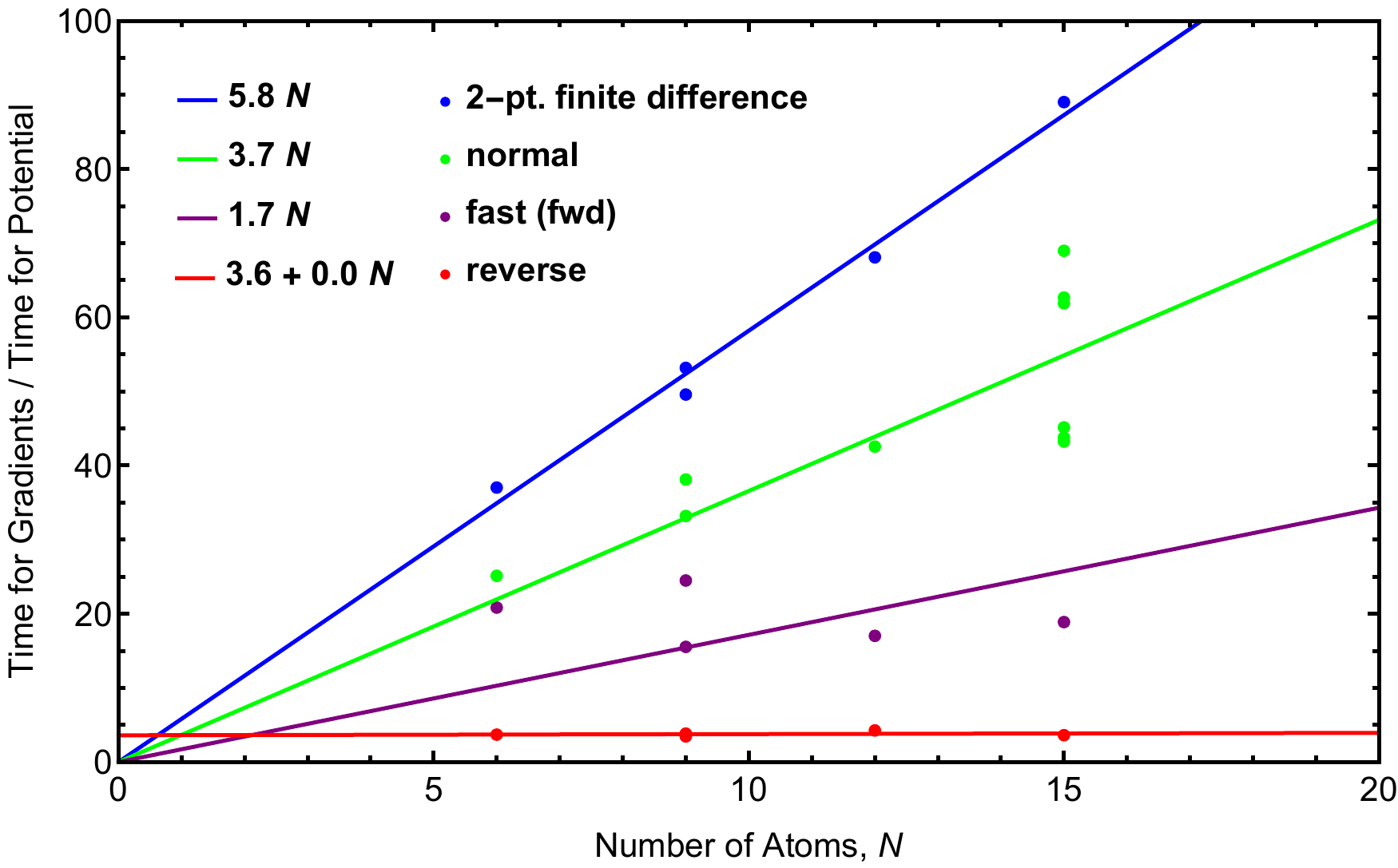}
    \caption{Calculation time for gradients, relative to that for the potential, for various methods as a function of the number of atoms, $N$.}
    \label{fig:timingplot}
\end{figure}

\subsection*{PIP Fast Gradient For Larger Molecules}
One might still question whether the advances in computer efficiency made possible by the reverse automated derivative method will stand up for systems other than ethanol and the 4-body water potential.  We have examined this question by comparing results from the water 2-body potential (6 atoms, unpublished), the ethanol potential (9 atoms, PIP orders 3 and 4, described above), the water 4-body potential (12 atoms),\cite{4body} the acetylacetone potential (15 atoms),\cite{QuAcAc} and the tropolone potential (15 atoms).\cite{tropolone20} The results for the timing cost for gradients, divided by that for the energy, are shown in Fig. \ref{fig:timingplot} for four different methods of differentiation: 2-point finite difference, normal analytical differentiation, fast (forward) differentiation, and reverse differentiation.  The last two methods have been described in this Communication. First-order fit parameters are shown in the legend; the first three are constrained to have zero intercept, while the reverse data is fit to an intercept and slope.  We noted earlier that the reverse method is predicted to have a nearly constant time cost, relative to the energy, of about 3-4.  The figure shows this to be substantially true for the number of atoms, $N$, between 6 and 15; there is negligible slope and the intercept is 3.6.  Scaling of the other methods is roughly as expected. Because there are two calls to the energy for the 2-point finite difference method, we might expect this method to go as $2 \times 3 N$; we find it to scale as $5.8 N$.  The normal differentiation needs to be performed $3 N$ times, so one might expect the cost to vary as $3 N$; it appears to vary as $3.7 N$. The cost of the ``fast'' method should be somewhere between that of the normal analytical differentiation and that of the reverse method; the result is $1.7 N$.  The reverse method is by far the fastest, and this advantage grows linearly with $N$. It should be noted that the time for the energy calculation itself varies non-linearly with $N$, depending on the symmetry and order. It is roughly proportional to the sum of the number of polynomials and monomials.

\subsection*{New Ethanol PES}
The new DFT-based PES for ethanol was done in just several ``human-days'' of effort using the well-established  protocol in our group.  It was fit with the same large PIP basis used for the assessment of PESs based on the MD17 dataset.  Thus, we consider this new PES mostly an example of the ease with which PESs for a molecule with 9 atoms and two methyl rotors can be developed and used for quantum simulations.  The immediate plan is to correct this PES using the CCSD(T) $\Delta$-ML approach we reported recently for N-methyl acetamide\cite{nandidelta_21} and acetylacetone.\cite{QudeltaJPCL_21}. This new PES along with extensive analysis will be reported shortly. However, for possible use in testing ML methods the extensive B3LYP dataset is available for download at https://scholarblogs.emory.edu/bowman/potential-energy-surfaces/.

\section*{Summary and Conclusions}
We reported new software incorporating reverse differentiation to obtain the gradient of a potential energy surface fit using permutationally invariant polynomials (PIPs). We demonstrated the substantial speed-up using this method, versus previous methods, for our recent 4-body water interaction potential. Detailed examination of training-testing precision and timing for ethanol using the  MD17 database of energies and gradients against GP-SOAP, ANI, sGDML, PhysNet, pKREG, KRR, ACE, etc was given.  These methods were recently assessed in detail by Dral and co-workers.\cite{dralchemsci} PIPs bases with 1898, 8895 and 14 572 terms were considered.  Training on energies plus gradients for a datasize of 250 configurations, the smallest PIP basis has RMSEs for energies and forces that are close to those from GAP-SOAP and sGDML (which are the best of all the other ML methods). Prediction times for 20 000 energies plus gradients, however, are very different (accounting for small documented differences in the various Intel CPUs). Normalized so that PIP is 1.0, sGDML and GAP-SOAP are 45 and 1395, respectively. The timings for sGDML and GAP-SOAP increase with training size whereas those for this PIP basis do not. However, the eRMSEs for sGDML and GAP-SOAP decrease to a final value of 0.1 kcal-mol$^{-1}$ which is about half the eRMSE for this small PIP basis. However, the prediction times grow substantially for sGDML and GAP-SOAP such that the times relative to this small PIP basis are 886 and 5000, respectively.  Ultimately among all the non-PIP methods neural-network PhysNet and ANI methods achieves the lowest energy and force RMSEs, roughly 0.06 kcal~mol$^{-1}$ and 0.20 kcal~mol$^{-1}$~A$^{-1}$ , respectively, when trained on 10 000 configurations. The largest PIP basis of 14 572 achieves very similar RMSEs but runs faster by factors of 144 and 18 compared to PhysNet and ANI, respectively. The middle-sized PIP basis of 8895 runs roughly 26 percent faster than the large PIP basis and when trained on 10 000 energies and gradients at 10 000 configurations achieving a testing energy and force RMSE of and 0.09 kcal~mol$^{-1}$  and 0.34 kcal~mol$^{-1}$~\AA$^{-1}$. 

Diffusion Monte Carlo calculations of the zero-point energy fail on the largest basis PIP PES trained on  MD17 dataset due to many ``holes''.   This is explained by noting that the energies of this dataset extend to only about 60\% of the ZPE.  A new  ethanol PIP PES is reported using  a B3LYP dataset of energies and gradients that extend to roughly 92 kcal~mol$^{-1}$. DMC calculations are successful using this PES. 

\section*{Supplementary Material}
The supplementary material contains examples of the reverse differentiation for PIP bases for a diatomic and a triatomic molecule, as well as Mathematica code for calculating an adjoint and a brief description of our Mathematica suite of programs. A comparison of the new ethanol PES and direct B3LYP optimized geometries of the minimum and normal mode frequencies is also given.

\section*{Acknowledgements}
JMB thanks NASA (80NSSC20K0360) for financial support. We thank Pavlo Dral for useful discussions and providing new plots of the learning curves including the new results using the PIP bases and also for timing using our PIP basis on same intel used to time the ML methods in ref. \cite{dralchemsci}.  We thank David Kovacs and Gabor Csanyi for sending details of the ACE timings and eRMSE values  and we thank Kurt Brorsen for running the SchNet timing calculations for \ce{OCHCO+}.

\section*{Data Availability}
The data that support the findings of this study are available from the corresponding author upon reasonable request. The B3LYP ethanol dataset is available for download at https://scholarblogs.emory.edu/bowman/potential-energy-surfaces/. The Mathematica notebooks used in this work are also available upon request.

\section*{References}
\bibliography{refs.bib} 

\begin{thebibliography}{7}%
\makeatletter
\providecommand \@ifxundefined [1]{%
 \@ifx{#1\undefined}
}%
\providecommand \@ifnum [1]{%
 \ifnum #1\expandafter \@firstoftwo
 \else \expandafter \@secondoftwo
 \fi
}%
\providecommand \@ifx [1]{%
 \ifx #1\expandafter \@firstoftwo
 \else \expandafter \@secondoftwo
 \fi
}%
\providecommand \natexlab [1]{#1}%
\providecommand \enquote  [1]{``#1''}%
\providecommand \bibnamefont  [1]{#1}%
\providecommand \bibfnamefont [1]{#1}%
\providecommand \citenamefont [1]{#1}%
\providecommand \href@noop [0]{\@secondoftwo}%
\providecommand \href [0]{\begingroup \@sanitize@url \@href}%
\providecommand \@href[1]{\@@startlink{#1}\@@href}%
\providecommand \@@href[1]{\endgroup#1\@@endlink}%
\providecommand \@sanitize@url [0]{\catcode `\\12\catcode `\$12\catcode
  `\&12\catcode `\#12\catcode `\^12\catcode `\_12\catcode `\%12\relax}%
\providecommand \@@startlink[1]{}%
\providecommand \@@endlink[0]{}%
\providecommand \url  [0]{\begingroup\@sanitize@url \@url }%
\providecommand \@url [1]{\endgroup\@href {#1}{\urlprefix }}%
\providecommand \urlprefix  [0]{URL }%
\providecommand \Eprint [0]{\href }%
\providecommand \doibase [0]{http://dx.doi.org/}%
\providecommand \selectlanguage [0]{\@gobble}%
\providecommand \bibinfo  [0]{\@secondoftwo}%
\providecommand \bibfield  [0]{\@secondoftwo}%
\providecommand \translation [1]{[#1]}%
\providecommand \BibitemOpen [0]{}%
\providecommand \bibitemStop [0]{}%
\providecommand \bibitemNoStop [0]{.\EOS\space}%
\providecommand \EOS [0]{\spacefactor3000\relax}%
\providecommand \BibitemShut  [1]{\csname bibitem#1\endcsname}%
\let\auto@bib@innerbib\@empty
\bibitem [{\citenamefont
  {Wolfram\hspace{.1cm}Research\hspace{.1cm}Inc.}(2019)}]{Mathematica}%
  \BibitemOpen
  \bibfield  {author} {\bibinfo {author} {\bibnamefont
  {Wolfram\hspace{.1cm}Research\hspace{.1cm}Inc.}},\ }\href@noop {} {\enquote
  {\bibinfo {title} {Mathematica, {V}ersion 12.0},}\ } (\bibinfo {year}
  {2019}),\ \bibinfo {note} {champaign, IL, 2019}\BibitemShut {NoStop}%
\bibitem [{Xie(2019)}]{Xie-nma}%
  \BibitemOpen
  \href {https://www.mcs.anl.gov/research/projects/msa/} {\enquote {\bibinfo
  {title} {Original msa software},}\ }\bibinfo {howpublished}
  {\url{https://www.mcs.anl.gov/research/projects/msa/}} (\bibinfo {year}
  {2019}),\ \bibinfo {note} {accessed: 2019-12-20}\BibitemShut {NoStop}%
\bibitem [{msa(2019)}]{msachen}%
  \BibitemOpen
  \href@noop {} {\enquote {\bibinfo {title} {Msa software with gradients},}\
  }\bibinfo {howpublished} {\url{https://github.com/szquchen/MSA-2.0}}
  (\bibinfo {year} {2019}),\ \bibinfo {note} {accessed: 2019-01-20}\BibitemShut
  {NoStop}%
\bibitem [{\citenamefont {Xie}\ and\ \citenamefont {Bowman}(2010)}]{Xie10}%
  \BibitemOpen
  \bibfield  {author} {\bibinfo {author} {\bibfnamefont {Z.}~\bibnamefont
  {Xie}}\ and\ \bibinfo {author} {\bibfnamefont {J.~M.}\ \bibnamefont
  {Bowman}},\ }\href {\doibase 10.1021/ct9004917} {\bibfield  {journal}
  {\bibinfo  {journal} {J. Chem. Theory Comput.}\ }\textbf {\bibinfo {volume}
  {6}},\ \bibinfo {pages} {26} (\bibinfo {year} {2010})}\BibitemShut {NoStop}%
\bibitem [{\citenamefont {Bowman}\ \emph {et~al.}(2010)\citenamefont {Bowman},
  \citenamefont {Braams}, \citenamefont {Carter}, \citenamefont {Chen},
  \citenamefont {Czakó}, \citenamefont {Fu}, \citenamefont {Huang},
  \citenamefont {Kamarchik}, \citenamefont {Sharma}, \citenamefont {Shepler},
  \citenamefont {Wang},\ and\ \citenamefont {Xie}}]{persp9}%
  \BibitemOpen
  \bibfield  {author} {\bibinfo {author} {\bibfnamefont {J.~M.}\ \bibnamefont
  {Bowman}}, \bibinfo {author} {\bibfnamefont {B.~J.}\ \bibnamefont {Braams}},
  \bibinfo {author} {\bibfnamefont {S.}~\bibnamefont {Carter}}, \bibinfo
  {author} {\bibfnamefont {C.}~\bibnamefont {Chen}}, \bibinfo {author}
  {\bibfnamefont {G.}~\bibnamefont {Czakó}}, \bibinfo {author} {\bibfnamefont
  {B.}~\bibnamefont {Fu}}, \bibinfo {author} {\bibfnamefont {X.}~\bibnamefont
  {Huang}}, \bibinfo {author} {\bibfnamefont {E.}~\bibnamefont {Kamarchik}},
  \bibinfo {author} {\bibfnamefont {A.~R.}\ \bibnamefont {Sharma}}, \bibinfo
  {author} {\bibfnamefont {B.~C.}\ \bibnamefont {Shepler}}, \bibinfo {author}
  {\bibfnamefont {Y.}~\bibnamefont {Wang}}, \ and\ \bibinfo {author}
  {\bibfnamefont {Z.}~\bibnamefont {Xie}},\ }\href {\doibase 10.1021/jz100626h}
  {\bibfield  {journal} {\bibinfo  {journal} {J. Phys. Chem. Lett.}\ }\textbf
  {\bibinfo {volume} {1}},\ \bibinfo {pages} {1866} (\bibinfo {year}
  {2010})}\BibitemShut {NoStop}%
\bibitem [{\citenamefont {Nandi}\ \emph {et~al.}(2019)\citenamefont {Nandi},
  \citenamefont {Qu},\ and\ \citenamefont {Bowman}}]{NandiBowman2019}%
  \BibitemOpen
  \bibfield  {author} {\bibinfo {author} {\bibfnamefont {A.}~\bibnamefont
  {Nandi}}, \bibinfo {author} {\bibfnamefont {C.}~\bibnamefont {Qu}}, \ and\
  \bibinfo {author} {\bibfnamefont {J.~M.}\ \bibnamefont {Bowman}},\
  }\href@noop {} {\bibfield  {journal} {\bibinfo  {journal} {J. Chem. Phys.}\
  }\textbf {\bibinfo {volume} {151}},\ \bibinfo {pages} {084306} (\bibinfo
  {year} {2019})}\BibitemShut {NoStop}%
\bibitem [{\citenamefont {Conte}\ \emph {et~al.}(2020)\citenamefont {Conte},
  \citenamefont {Qu}, \citenamefont {Houston},\ and\ \citenamefont
  {Bowman}}]{conte20_efficientPIP}%
  \BibitemOpen
  \bibfield  {author} {\bibinfo {author} {\bibfnamefont {R.}~\bibnamefont
  {Conte}}, \bibinfo {author} {\bibfnamefont {C.}~\bibnamefont {Qu}}, \bibinfo
  {author} {\bibfnamefont {P.~L.}\ \bibnamefont {Houston}}, \ and\ \bibinfo
  {author} {\bibfnamefont {J.~M.}\ \bibnamefont {Bowman}},\ }\href@noop {}
  {\bibfield  {journal} {\bibinfo  {journal} {J. Chem. Theory Comput.}\
  }\textbf {\bibinfo {volume} {16}},\ \bibinfo {pages} {3264} (\bibinfo {year}
  {2020})}\BibitemShut {NoStop}%
\end{thebibliography}%


\begin{thebibliography}{70}%
\makeatletter
\providecommand \@ifxundefined [1]{%
 \@ifx{#1\undefined}
}%
\providecommand \@ifnum [1]{%
 \ifnum #1\expandafter \@firstoftwo
 \else \expandafter \@secondoftwo
 \fi
}%
\providecommand \@ifx [1]{%
 \ifx #1\expandafter \@firstoftwo
 \else \expandafter \@secondoftwo
 \fi
}%
\providecommand \natexlab [1]{#1}%
\providecommand \enquote  [1]{``#1''}%
\providecommand \bibnamefont  [1]{#1}%
\providecommand \bibfnamefont [1]{#1}%
\providecommand \citenamefont [1]{#1}%
\providecommand \href@noop [0]{\@secondoftwo}%
\providecommand \href [0]{\begingroup \@sanitize@url \@href}%
\providecommand \@href[1]{\@@startlink{#1}\@@href}%
\providecommand \@@href[1]{\endgroup#1\@@endlink}%
\providecommand \@sanitize@url [0]{\catcode `\\12\catcode `\$12\catcode
  `\&12\catcode `\#12\catcode `\^12\catcode `\_12\catcode `\%12\relax}%
\providecommand \@@startlink[1]{}%
\providecommand \@@endlink[0]{}%
\providecommand \url  [0]{\begingroup\@sanitize@url \@url }%
\providecommand \@url [1]{\endgroup\@href {#1}{\urlprefix }}%
\providecommand \urlprefix  [0]{URL }%
\providecommand \Eprint [0]{\href }%
\providecommand \doibase [0]{http://dx.doi.org/}%
\providecommand \selectlanguage [0]{\@gobble}%
\providecommand \bibinfo  [0]{\@secondoftwo}%
\providecommand \bibfield  [0]{\@secondoftwo}%
\providecommand \translation [1]{[#1]}%
\providecommand \BibitemOpen [0]{}%
\providecommand \bibitemStop [0]{}%
\providecommand \bibitemNoStop [0]{.\EOS\space}%
\providecommand \EOS [0]{\spacefactor3000\relax}%
\providecommand \BibitemShut  [1]{\csname bibitem#1\endcsname}%
\let\auto@bib@innerbib\@empty
\bibitem [{\citenamefont {Westermayr}\ \emph {et~al.}(2021)\citenamefont
  {Westermayr}, \citenamefont {Gastegger}, \citenamefont {Schütt},\ and\
  \citenamefont {Maurer}}]{persp1}%
  \BibitemOpen
  \bibfield  {author} {\bibinfo {author} {\bibfnamefont {J.}~\bibnamefont
  {Westermayr}}, \bibinfo {author} {\bibfnamefont {M.}~\bibnamefont
  {Gastegger}}, \bibinfo {author} {\bibfnamefont {K.~T.}\ \bibnamefont
  {Schütt}}, \ and\ \bibinfo {author} {\bibfnamefont {R.~J.}\ \bibnamefont
  {Maurer}},\ }\href {\doibase 10.1063/5.0047760} {\bibfield  {journal}
  {\bibinfo  {journal} {J. Chem. Phys.}\ }\textbf {\bibinfo {volume} {154}},\
  \bibinfo {pages} {230903} (\bibinfo {year} {2021})}\BibitemShut {NoStop}%
\bibitem [{\citenamefont {Koner}\ \emph {et~al.}(2020)\citenamefont {Koner},
  \citenamefont {Salehi}, \citenamefont {Mondal},\ and\ \citenamefont
  {Meuwly}}]{persp2}%
  \BibitemOpen
  \bibfield  {author} {\bibinfo {author} {\bibfnamefont {D.}~\bibnamefont
  {Koner}}, \bibinfo {author} {\bibfnamefont {S.~M.}\ \bibnamefont {Salehi}},
  \bibinfo {author} {\bibfnamefont {P.}~\bibnamefont {Mondal}}, \ and\ \bibinfo
  {author} {\bibfnamefont {M.}~\bibnamefont {Meuwly}},\ }\href {\doibase
  10.1063/5.0009628} {\bibfield  {journal} {\bibinfo  {journal} {J. Chem.
  Phys.}\ }\textbf {\bibinfo {volume} {153}},\ \bibinfo {pages} {010901}
  (\bibinfo {year} {2020})}\BibitemShut {NoStop}%
\bibitem [{\citenamefont {Fröhlking}\ \emph {et~al.}(2020)\citenamefont
  {Fröhlking}, \citenamefont {Bernetti}, \citenamefont {Calonaci},\ and\
  \citenamefont {Bussi}}]{persp3}%
  \BibitemOpen
  \bibfield  {author} {\bibinfo {author} {\bibfnamefont {T.}~\bibnamefont
  {Fröhlking}}, \bibinfo {author} {\bibfnamefont {M.}~\bibnamefont
  {Bernetti}}, \bibinfo {author} {\bibfnamefont {N.}~\bibnamefont {Calonaci}},
  \ and\ \bibinfo {author} {\bibfnamefont {G.}~\bibnamefont {Bussi}},\ }\href
  {\doibase 10.1063/5.0011346} {\bibfield  {journal} {\bibinfo  {journal} {J.
  Chem. Phys.}\ }\textbf {\bibinfo {volume} {152}},\ \bibinfo {pages} {230902}
  (\bibinfo {year} {2020})}\BibitemShut {NoStop}%
\bibitem [{\citenamefont {Tong}\ \emph {et~al.}(2020)\citenamefont {Tong},
  \citenamefont {Gao}, \citenamefont {Liu}, \citenamefont {Xie}, \citenamefont
  {Lv}, \citenamefont {Wang},\ and\ \citenamefont {Zhao}}]{perps4}%
  \BibitemOpen
  \bibfield  {author} {\bibinfo {author} {\bibfnamefont {Q.}~\bibnamefont
  {Tong}}, \bibinfo {author} {\bibfnamefont {P.}~\bibnamefont {Gao}}, \bibinfo
  {author} {\bibfnamefont {H.}~\bibnamefont {Liu}}, \bibinfo {author}
  {\bibfnamefont {Y.}~\bibnamefont {Xie}}, \bibinfo {author} {\bibfnamefont
  {J.}~\bibnamefont {Lv}}, \bibinfo {author} {\bibfnamefont {Y.}~\bibnamefont
  {Wang}}, \ and\ \bibinfo {author} {\bibfnamefont {J.}~\bibnamefont {Zhao}},\
  }\href {\doibase 10.1021/acs.jpclett.0c02357} {\bibfield  {journal} {\bibinfo
   {journal} {J. Phys. Chem. Lett.}\ }\textbf {\bibinfo {volume} {11}},\
  \bibinfo {pages} {8710} (\bibinfo {year} {2020})}\BibitemShut {NoStop}%
\bibitem [{\citenamefont {Jinnouchi}\ \emph {et~al.}(2020)\citenamefont
  {Jinnouchi}, \citenamefont {Miwa}, \citenamefont {Karsai}, \citenamefont
  {Kresse},\ and\ \citenamefont {Asahi}}]{persp5}%
  \BibitemOpen
  \bibfield  {author} {\bibinfo {author} {\bibfnamefont {R.}~\bibnamefont
  {Jinnouchi}}, \bibinfo {author} {\bibfnamefont {K.}~\bibnamefont {Miwa}},
  \bibinfo {author} {\bibfnamefont {F.}~\bibnamefont {Karsai}}, \bibinfo
  {author} {\bibfnamefont {G.}~\bibnamefont {Kresse}}, \ and\ \bibinfo {author}
  {\bibfnamefont {R.}~\bibnamefont {Asahi}},\ }\href {\doibase
  10.1021/acs.jpclett.0c01061} {\bibfield  {journal} {\bibinfo  {journal} {J.
  Phys. Chem. Lett.}\ }\textbf {\bibinfo {volume} {11}},\ \bibinfo {pages}
  {6946} (\bibinfo {year} {2020})}\BibitemShut {NoStop}%
\bibitem [{\citenamefont {Jiang}, \citenamefont {Li},\ and\ \citenamefont
  {Guo}(2020)}]{persp6}%
  \BibitemOpen
  \bibfield  {author} {\bibinfo {author} {\bibfnamefont {B.}~\bibnamefont
  {Jiang}}, \bibinfo {author} {\bibfnamefont {J.}~\bibnamefont {Li}}, \ and\
  \bibinfo {author} {\bibfnamefont {H.}~\bibnamefont {Guo}},\ }\href {\doibase
  10.1021/acs.jpclett.0c00989} {\bibfield  {journal} {\bibinfo  {journal} {J.
  Phys. Chem. Lett.}\ }\textbf {\bibinfo {volume} {11}},\ \bibinfo {pages}
  {5120} (\bibinfo {year} {2020})}\BibitemShut {NoStop}%
\bibitem [{\citenamefont {Dral}(2020)}]{persp7}%
  \BibitemOpen
  \bibfield  {author} {\bibinfo {author} {\bibfnamefont {P.~O.}\ \bibnamefont
  {Dral}},\ }\href {\doibase 10.1021/acs.jpclett.9b03664} {\bibfield  {journal}
  {\bibinfo  {journal} {J. Phys. Chem. Lett.}\ }\textbf {\bibinfo {volume}
  {11}},\ \bibinfo {pages} {2336} (\bibinfo {year} {2020})}\BibitemShut
  {NoStop}%
\bibitem [{\citenamefont {Poltavsky}\ and\ \citenamefont
  {Tkatchenko}(2021)}]{persp8}%
  \BibitemOpen
  \bibfield  {author} {\bibinfo {author} {\bibfnamefont {I.}~\bibnamefont
  {Poltavsky}}\ and\ \bibinfo {author} {\bibfnamefont {A.}~\bibnamefont
  {Tkatchenko}},\ }\href {\doibase 10.1021/acs.jpclett.1c01204} {\bibfield
  {journal} {\bibinfo  {journal} {J. Phys. Chem. Lett.}\ }\textbf {\bibinfo
  {volume} {12}},\ \bibinfo {pages} {6551} (\bibinfo {year}
  {2021})}\BibitemShut {NoStop}%
\bibitem [{\citenamefont {Bowman}\ \emph {et~al.}(2010)\citenamefont {Bowman},
  \citenamefont {Braams}, \citenamefont {Carter}, \citenamefont {Chen},
  \citenamefont {Czakó}, \citenamefont {Fu}, \citenamefont {Huang},
  \citenamefont {Kamarchik}, \citenamefont {Sharma}, \citenamefont {Shepler},
  \citenamefont {Wang},\ and\ \citenamefont {Xie}}]{persp9}%
  \BibitemOpen
  \bibfield  {author} {\bibinfo {author} {\bibfnamefont {J.~M.}\ \bibnamefont
  {Bowman}}, \bibinfo {author} {\bibfnamefont {B.~J.}\ \bibnamefont {Braams}},
  \bibinfo {author} {\bibfnamefont {S.}~\bibnamefont {Carter}}, \bibinfo
  {author} {\bibfnamefont {C.}~\bibnamefont {Chen}}, \bibinfo {author}
  {\bibfnamefont {G.}~\bibnamefont {Czakó}}, \bibinfo {author} {\bibfnamefont
  {B.}~\bibnamefont {Fu}}, \bibinfo {author} {\bibfnamefont {X.}~\bibnamefont
  {Huang}}, \bibinfo {author} {\bibfnamefont {E.}~\bibnamefont {Kamarchik}},
  \bibinfo {author} {\bibfnamefont {A.~R.}\ \bibnamefont {Sharma}}, \bibinfo
  {author} {\bibfnamefont {B.~C.}\ \bibnamefont {Shepler}}, \bibinfo {author}
  {\bibfnamefont {Y.}~\bibnamefont {Wang}}, \ and\ \bibinfo {author}
  {\bibfnamefont {Z.}~\bibnamefont {Xie}},\ }\href {\doibase 10.1021/jz100626h}
  {\bibfield  {journal} {\bibinfo  {journal} {J. Phys. Chem. Lett.}\ }\textbf
  {\bibinfo {volume} {1}},\ \bibinfo {pages} {1866} (\bibinfo {year}
  {2010})}\BibitemShut {NoStop}%
\bibitem [{\citenamefont {Brown}\ \emph {et~al.}(2003)\citenamefont {Brown},
  \citenamefont {Braams}, \citenamefont {Christoffel}, \citenamefont {Jin},\
  and\ \citenamefont {Bowman}}]{Bowman2003}%
  \BibitemOpen
  \bibfield  {author} {\bibinfo {author} {\bibfnamefont {A.}~\bibnamefont
  {Brown}}, \bibinfo {author} {\bibfnamefont {B.~J.}\ \bibnamefont {Braams}},
  \bibinfo {author} {\bibfnamefont {K.}~\bibnamefont {Christoffel}}, \bibinfo
  {author} {\bibfnamefont {Z.}~\bibnamefont {Jin}}, \ and\ \bibinfo {author}
  {\bibfnamefont {J.~M.}\ \bibnamefont {Bowman}},\ }\href {\doibase
  10.1063/1.1622379} {\bibfield  {journal} {\bibinfo  {journal} {J. Chem.
  Phys.}\ }\textbf {\bibinfo {volume} {119}},\ \bibinfo {pages} {8790}
  (\bibinfo {year} {2003})}\BibitemShut {NoStop}%
\bibitem [{\citenamefont {Qu}, \citenamefont {Yu},\ and\ \citenamefont
  {Bowman}(2018)}]{ARPC2018}%
  \BibitemOpen
  \bibfield  {author} {\bibinfo {author} {\bibfnamefont {C.}~\bibnamefont
  {Qu}}, \bibinfo {author} {\bibfnamefont {Q.}~\bibnamefont {Yu}}, \ and\
  \bibinfo {author} {\bibfnamefont {J.~M.}\ \bibnamefont {Bowman}},\ }\href
  {\doibase 10.1146/annurev-physchem-050317-021139} {\bibfield  {journal}
  {\bibinfo  {journal} {Annu. Rev. Phys. Chem.}\ }\textbf {\bibinfo {volume}
  {69}},\ \bibinfo {pages} {6.1} (\bibinfo {year} {2018})}\BibitemShut
  {NoStop}%
\bibitem [{\citenamefont {Győri}\ and\ \citenamefont {Czakó}(2020)}]{robo20}%
  \BibitemOpen
  \bibfield  {author} {\bibinfo {author} {\bibfnamefont {T.}~\bibnamefont
  {Győri}}\ and\ \bibinfo {author} {\bibfnamefont {G.}~\bibnamefont
  {Czakó}},\ }\href {\doibase 10.1021/acs.jctc.9b01006} {\bibfield  {journal}
  {\bibinfo  {journal} {J. Comput. Theory Chem.}\ }\textbf {\bibinfo {volume}
  {16}},\ \bibinfo {pages} {51} (\bibinfo {year} {2020})},\ \bibinfo {note}
  {pMID: 31851508}\BibitemShut {NoStop}%
\bibitem [{\citenamefont {Reddy}\ \emph {et~al.}(2016)\citenamefont {Reddy},
  \citenamefont {Straight}, \citenamefont {Bajaj}, \citenamefont {Huy~Pham},
  \citenamefont {Riera}, \citenamefont {Moberg}, \citenamefont {Morales},
  \citenamefont {Knight}, \citenamefont {Götz},\ and\ \citenamefont
  {Paesani}}]{mbpoltests}%
  \BibitemOpen
  \bibfield  {author} {\bibinfo {author} {\bibfnamefont {S.~K.}\ \bibnamefont
  {Reddy}}, \bibinfo {author} {\bibfnamefont {S.~C.}\ \bibnamefont {Straight}},
  \bibinfo {author} {\bibfnamefont {P.}~\bibnamefont {Bajaj}}, \bibinfo
  {author} {\bibfnamefont {C.}~\bibnamefont {Huy~Pham}}, \bibinfo {author}
  {\bibfnamefont {M.}~\bibnamefont {Riera}}, \bibinfo {author} {\bibfnamefont
  {D.~R.}\ \bibnamefont {Moberg}}, \bibinfo {author} {\bibfnamefont {M.~A.}\
  \bibnamefont {Morales}}, \bibinfo {author} {\bibfnamefont {C.}~\bibnamefont
  {Knight}}, \bibinfo {author} {\bibfnamefont {A.~W.}\ \bibnamefont {Götz}}, \
  and\ \bibinfo {author} {\bibfnamefont {F.}~\bibnamefont {Paesani}},\ }\href
  {\doibase 10.1063/1.4967719} {\bibfield  {journal} {\bibinfo  {journal} {J.
  Chem. Phys.}\ }\textbf {\bibinfo {volume} {145}},\ \bibinfo {pages} {194504}
  (\bibinfo {year} {2016})}\BibitemShut {NoStop}%
\bibitem [{\citenamefont {Lambros}\ \emph {et~al.}(2021)\citenamefont
  {Lambros}, \citenamefont {Dasgupta}, \citenamefont {Palos}, \citenamefont
  {Swee}, \citenamefont {Hu},\ and\ \citenamefont {Paesani}}]{paespips_21}%
  \BibitemOpen
  \bibfield  {author} {\bibinfo {author} {\bibfnamefont {E.}~\bibnamefont
  {Lambros}}, \bibinfo {author} {\bibfnamefont {S.}~\bibnamefont {Dasgupta}},
  \bibinfo {author} {\bibfnamefont {E.}~\bibnamefont {Palos}}, \bibinfo
  {author} {\bibfnamefont {S.}~\bibnamefont {Swee}}, \bibinfo {author}
  {\bibfnamefont {J.}~\bibnamefont {Hu}}, \ and\ \bibinfo {author}
  {\bibfnamefont {F.}~\bibnamefont {Paesani}},\ }\href {\doibase
  10.1021/acs.jctc.1c00541} {\bibfield  {journal} {\bibinfo  {journal} {J.
  Chem. Theory Comput.}\ }\textbf {\bibinfo {volume} {17}},\ \bibinfo {pages}
  {5635} (\bibinfo {year} {2021})}\BibitemShut {NoStop}%
\bibitem [{\citenamefont {Conte}\ \emph
  {et~al.}(2020{\natexlab{a}})\citenamefont {Conte}, \citenamefont {Qu},
  \citenamefont {Houston},\ and\ \citenamefont
  {Bowman}}]{conte20_efficientPIP}%
  \BibitemOpen
  \bibfield  {author} {\bibinfo {author} {\bibfnamefont {R.}~\bibnamefont
  {Conte}}, \bibinfo {author} {\bibfnamefont {C.}~\bibnamefont {Qu}}, \bibinfo
  {author} {\bibfnamefont {P.~L.}\ \bibnamefont {Houston}}, \ and\ \bibinfo
  {author} {\bibfnamefont {J.~M.}\ \bibnamefont {Bowman}},\ }\href@noop {}
  {\bibfield  {journal} {\bibinfo  {journal} {J. Chem. Theory Comput.}\
  }\textbf {\bibinfo {volume} {16}},\ \bibinfo {pages} {3264} (\bibinfo {year}
  {2020}{\natexlab{a}})}\BibitemShut {NoStop}%
\bibitem [{\citenamefont {Moberg}\ and\ \citenamefont
  {Jasper}(2021)}]{jasperpips_21}%
  \BibitemOpen
  \bibfield  {author} {\bibinfo {author} {\bibfnamefont {D.~R.}\ \bibnamefont
  {Moberg}}\ and\ \bibinfo {author} {\bibfnamefont {A.~W.}\ \bibnamefont
  {Jasper}},\ }\href {\doibase 10.1021/acs.jctc.1c00352} {\bibfield  {journal}
  {\bibinfo  {journal} {Journal of Chemical Theory and Computation}\ }\textbf
  {\bibinfo {volume} {17}},\ \bibinfo {pages} {5440} (\bibinfo {year}
  {2021})}\BibitemShut {NoStop}%
\bibitem [{\citenamefont {Moberg}, \citenamefont {Jasper},\ and\ \citenamefont
  {Davis}(2021)}]{greedypip_21}%
  \BibitemOpen
  \bibfield  {author} {\bibinfo {author} {\bibfnamefont {D.~R.}\ \bibnamefont
  {Moberg}}, \bibinfo {author} {\bibfnamefont {A.~W.}\ \bibnamefont {Jasper}},
  \ and\ \bibinfo {author} {\bibfnamefont {M.~J.}\ \bibnamefont {Davis}},\
  }\href {\doibase 10.1021/acs.jpclett.1c02721} {\bibfield  {journal} {\bibinfo
   {journal} {J. Phys. Chem. Lett.}\ }\textbf {\bibinfo {volume} {12}},\
  \bibinfo {pages} {9169} (\bibinfo {year} {2021})},\ \bibinfo {note} {pMID:
  34525799}\BibitemShut {NoStop}%
\bibitem [{\citenamefont {Jiang}, \citenamefont {Li},\ and\ \citenamefont
  {Guo}(2016)}]{Guo16}%
  \BibitemOpen
  \bibfield  {author} {\bibinfo {author} {\bibfnamefont {B.}~\bibnamefont
  {Jiang}}, \bibinfo {author} {\bibfnamefont {J.}~\bibnamefont {Li}}, \ and\
  \bibinfo {author} {\bibfnamefont {H.}~\bibnamefont {Guo}},\ }\href {\doibase
  10.1080/0144235X.2016.1200347} {\bibfield  {journal} {\bibinfo  {journal}
  {Int. Rev. Phys. Chem.}\ }\textbf {\bibinfo {volume} {35}},\ \bibinfo {pages}
  {479} (\bibinfo {year} {2016})}\BibitemShut {NoStop}%
\bibitem [{\citenamefont {Shao}\ \emph {et~al.}(2016)\citenamefont {Shao},
  \citenamefont {Chen}, \citenamefont {Zhao},\ and\ \citenamefont
  {Zhang}}]{NNZhang16}%
  \BibitemOpen
  \bibfield  {author} {\bibinfo {author} {\bibfnamefont {K.}~\bibnamefont
  {Shao}}, \bibinfo {author} {\bibfnamefont {J.}~\bibnamefont {Chen}}, \bibinfo
  {author} {\bibfnamefont {Z.}~\bibnamefont {Zhao}}, \ and\ \bibinfo {author}
  {\bibfnamefont {D.~H.}\ \bibnamefont {Zhang}},\ }\href {\doibase
  10.1063/1.4961454} {\bibfield  {journal} {\bibinfo  {journal} {J. Chem.
  Phys.}\ }\textbf {\bibinfo {volume} {145}},\ \bibinfo {pages} {071101}
  (\bibinfo {year} {2016})}\BibitemShut {NoStop}%
\bibitem [{\citenamefont {Fu}\ and\ \citenamefont {Zhang}(2018)}]{FINN_18}%
  \BibitemOpen
  \bibfield  {author} {\bibinfo {author} {\bibfnamefont {B.}~\bibnamefont
  {Fu}}\ and\ \bibinfo {author} {\bibfnamefont {D.~H.}\ \bibnamefont {Zhang}},\
  }\href {\doibase 10.1021/acs.jctc.8b00006} {\bibfield  {journal} {\bibinfo
  {journal} {J. Chem. Theory Comput.}\ }\textbf {\bibinfo {volume} {14}},\
  \bibinfo {pages} {2289} (\bibinfo {year} {2018})}\BibitemShut {NoStop}%
\bibitem [{\citenamefont {Qu}\ \emph {et~al.}(2018)\citenamefont {Qu},
  \citenamefont {Yu}, \citenamefont {Van~Hoozen}, \citenamefont {Bowman},\ and\
  \citenamefont {Vargas-Hern{\'a}ndez}}]{PIP-GP}%
  \BibitemOpen
  \bibfield  {author} {\bibinfo {author} {\bibfnamefont {C.}~\bibnamefont
  {Qu}}, \bibinfo {author} {\bibfnamefont {Q.}~\bibnamefont {Yu}}, \bibinfo
  {author} {\bibfnamefont {B.~L.}\ \bibnamefont {Van~Hoozen}}, \bibinfo
  {author} {\bibfnamefont {J.~M.}\ \bibnamefont {Bowman}}, \ and\ \bibinfo
  {author} {\bibfnamefont {R.~A.}\ \bibnamefont {Vargas-Hern{\'a}ndez}},\
  }\href {\doibase 10.1021/acs.jctc.8b00298} {\bibfield  {journal} {\bibinfo
  {journal} {J. Chem. Theory Comput.}\ }\textbf {\bibinfo {volume} {14}},\
  \bibinfo {pages} {3381} (\bibinfo {year} {2018})}\BibitemShut {NoStop}%
\bibitem [{\citenamefont {Allen}\ \emph {et~al.}(2021)\citenamefont {Allen},
  \citenamefont {Dusson}, \citenamefont {Ortner},\ and\ \citenamefont
  {Cs{\'{a}}nyi}}]{Xie10Allen2021}%
  \BibitemOpen
  \bibfield  {author} {\bibinfo {author} {\bibfnamefont {A.~E.~A.}\
  \bibnamefont {Allen}}, \bibinfo {author} {\bibfnamefont {G.}~\bibnamefont
  {Dusson}}, \bibinfo {author} {\bibfnamefont {C.}~\bibnamefont {Ortner}}, \
  and\ \bibinfo {author} {\bibfnamefont {G.}~\bibnamefont {Cs{\'{a}}nyi}},\
  }\href {\doibase 10.1088/2632-2153/abd51e} {\bibfield  {journal} {\bibinfo
  {journal} {Mach. Learn.: Sci. Technol.}\ }\textbf {\bibinfo {volume} {2}},\
  \bibinfo {pages} {025017} (\bibinfo {year} {2021})}\BibitemShut {NoStop}%
\bibitem [{\citenamefont {van~der Oord}\ \emph {et~al.}(2020)\citenamefont
  {van~der Oord}, \citenamefont {Dusson}, \citenamefont {Cs{\'{a}}nyi},\ and\
  \citenamefont {Ortner}}]{Oord2020}%
  \BibitemOpen
  \bibfield  {author} {\bibinfo {author} {\bibfnamefont {C.}~\bibnamefont
  {van~der Oord}}, \bibinfo {author} {\bibfnamefont {G.}~\bibnamefont
  {Dusson}}, \bibinfo {author} {\bibfnamefont {G.}~\bibnamefont
  {Cs{\'{a}}nyi}}, \ and\ \bibinfo {author} {\bibfnamefont {C.}~\bibnamefont
  {Ortner}},\ }\href {\doibase 10.1088/2632-2153/ab527c} {\bibfield  {journal}
  {\bibinfo  {journal} {Mach. Learn.: Sci. Technol.}\ }\textbf {\bibinfo
  {volume} {1}},\ \bibinfo {pages} {015004} (\bibinfo {year}
  {2020})}\BibitemShut {NoStop}%
\bibitem [{\citenamefont {Nguyen}\ \emph {et~al.}(2018)\citenamefont {Nguyen},
  \citenamefont {Székely}, \citenamefont {Imbalzano}, \citenamefont {Behler},
  \citenamefont {Csányi}, \citenamefont {Ceriotti}, \citenamefont {Götz},\
  and\ \citenamefont {Paesani}}]{paescomps}%
  \BibitemOpen
  \bibfield  {author} {\bibinfo {author} {\bibfnamefont {T.~T.}\ \bibnamefont
  {Nguyen}}, \bibinfo {author} {\bibfnamefont {E.}~\bibnamefont {Székely}},
  \bibinfo {author} {\bibfnamefont {G.}~\bibnamefont {Imbalzano}}, \bibinfo
  {author} {\bibfnamefont {J.}~\bibnamefont {Behler}}, \bibinfo {author}
  {\bibfnamefont {G.}~\bibnamefont {Csányi}}, \bibinfo {author} {\bibfnamefont
  {M.}~\bibnamefont {Ceriotti}}, \bibinfo {author} {\bibfnamefont {A.~W.}\
  \bibnamefont {Götz}}, \ and\ \bibinfo {author} {\bibfnamefont
  {F.}~\bibnamefont {Paesani}},\ }\href {\doibase 10.1063/1.5024577} {\bibfield
   {journal} {\bibinfo  {journal} {J. Chem. Phys.}\ }\textbf {\bibinfo {volume}
  {148}},\ \bibinfo {pages} {241725} (\bibinfo {year} {2018})}\BibitemShut
  {NoStop}%
\bibitem [{\citenamefont {Sauceda}\ \emph
  {et~al.}(2019{\natexlab{a}})\citenamefont {Sauceda}, \citenamefont {Chmiela},
  \citenamefont {Poltavsky}, \citenamefont {Müller},\ and\ \citenamefont
  {Tkatchenko}}]{Tkatchjcp}%
  \BibitemOpen
  \bibfield  {author} {\bibinfo {author} {\bibfnamefont {H.~E.}\ \bibnamefont
  {Sauceda}}, \bibinfo {author} {\bibfnamefont {S.}~\bibnamefont {Chmiela}},
  \bibinfo {author} {\bibfnamefont {I.}~\bibnamefont {Poltavsky}}, \bibinfo
  {author} {\bibfnamefont {K.-R.}\ \bibnamefont {Müller}}, \ and\ \bibinfo
  {author} {\bibfnamefont {A.}~\bibnamefont {Tkatchenko}},\ }\href@noop {}
  {\bibfield  {journal} {\bibinfo  {journal} {J. Chem. Phys}\ }\textbf
  {\bibinfo {volume} {150}},\ \bibinfo {pages} {114102} (\bibinfo {year}
  {2019}{\natexlab{a}})}\BibitemShut {NoStop}%
\bibitem [{\citenamefont {Pinheiro}\ \emph {et~al.}(2021)\citenamefont
  {Pinheiro}, \citenamefont {Ge}, \citenamefont {Ferré}, \citenamefont
  {Dral},\ and\ \citenamefont {Barbatti}}]{dralchemsci}%
  \BibitemOpen
  \bibfield  {author} {\bibinfo {author} {\bibfnamefont {M.}~\bibnamefont
  {Pinheiro}}, \bibinfo {author} {\bibfnamefont {F.}~\bibnamefont {Ge}},
  \bibinfo {author} {\bibfnamefont {N.}~\bibnamefont {Ferré}}, \bibinfo
  {author} {\bibfnamefont {P.~O.}\ \bibnamefont {Dral}}, \ and\ \bibinfo
  {author} {\bibfnamefont {M.}~\bibnamefont {Barbatti}},\ }\href {\doibase
  10.1039/D1SC03564A} {\bibfield  {journal} {\bibinfo  {journal} {Chem. Sci.}\
  }\textbf {\bibinfo {volume} {12}},\ \bibinfo {pages} {14396} (\bibinfo {year}
  {2021})}\BibitemShut {NoStop}%
\bibitem [{\citenamefont {Chmiela}\ \emph {et~al.}(2019)\citenamefont
  {Chmiela}, \citenamefont {Sauceda}, \citenamefont {Poltavsky}, \citenamefont
  {Müller},\ and\ \citenamefont {Tkatchenko}}]{MD17}%
  \BibitemOpen
  \bibfield  {author} {\bibinfo {author} {\bibfnamefont {S.}~\bibnamefont
  {Chmiela}}, \bibinfo {author} {\bibfnamefont {H.~E.}\ \bibnamefont
  {Sauceda}}, \bibinfo {author} {\bibfnamefont {I.}~\bibnamefont {Poltavsky}},
  \bibinfo {author} {\bibfnamefont {K.-R.}\ \bibnamefont {Müller}}, \ and\
  \bibinfo {author} {\bibfnamefont {A.}~\bibnamefont {Tkatchenko}},\ }\href
  {\doibase https://doi.org/10.1016/j.cpc.2019.02.007} {\bibfield  {journal}
  {\bibinfo  {journal} {Comp. Phys. Comm.}\ }\textbf {\bibinfo {volume}
  {240}},\ \bibinfo {pages} {38} (\bibinfo {year} {2019})}\BibitemShut
  {NoStop}%
\bibitem [{\citenamefont {Bart\'ok}\ and\ \citenamefont
  {Cs\'anyi}(2015)}]{GP-2015-1}%
  \BibitemOpen
  \bibfield  {author} {\bibinfo {author} {\bibfnamefont {A.~P.}\ \bibnamefont
  {Bart\'ok}}\ and\ \bibinfo {author} {\bibfnamefont {G.}~\bibnamefont
  {Cs\'anyi}},\ }\href {\doibase 10.1002/qua.24927} {\bibfield  {journal}
  {\bibinfo  {journal} {Int. J. Quantum Chem.}\ }\textbf {\bibinfo {volume}
  {115}},\ \bibinfo {pages} {1051} (\bibinfo {year} {2015})}\BibitemShut
  {NoStop}%
\bibitem [{\citenamefont {Smith}, \citenamefont {Isayev},\ and\ \citenamefont
  {Roitberg}(2017)}]{AN1}%
  \BibitemOpen
  \bibfield  {author} {\bibinfo {author} {\bibfnamefont {J.~S.}\ \bibnamefont
  {Smith}}, \bibinfo {author} {\bibfnamefont {O.}~\bibnamefont {Isayev}}, \
  and\ \bibinfo {author} {\bibfnamefont {A.~E.}\ \bibnamefont {Roitberg}},\
  }\href {\doibase 10.1039/C6SC05720A} {\bibfield  {journal} {\bibinfo
  {journal} {Chem. Sci.}\ }\textbf {\bibinfo {volume} {8}},\ \bibinfo {pages}
  {3192} (\bibinfo {year} {2017})}\BibitemShut {NoStop}%
\bibitem [{\citenamefont {Zhang}\ \emph {et~al.}(2018)\citenamefont {Zhang},
  \citenamefont {Han}, \citenamefont {Wang}, \citenamefont {Car},\ and\
  \citenamefont {E}}]{dpmd2018}%
  \BibitemOpen
  \bibfield  {author} {\bibinfo {author} {\bibfnamefont {L.}~\bibnamefont
  {Zhang}}, \bibinfo {author} {\bibfnamefont {J.}~\bibnamefont {Han}}, \bibinfo
  {author} {\bibfnamefont {H.}~\bibnamefont {Wang}}, \bibinfo {author}
  {\bibfnamefont {R.}~\bibnamefont {Car}}, \ and\ \bibinfo {author}
  {\bibfnamefont {W.}~\bibnamefont {E}},\ }\href {\doibase
  10.1103/PhysRevLett.120.143001} {\bibfield  {journal} {\bibinfo  {journal}
  {Phys. Rev. Lett.}\ }\textbf {\bibinfo {volume} {120}},\ \bibinfo {pages}
  {143001} (\bibinfo {year} {2018})}\BibitemShut {NoStop}%
\bibitem [{\citenamefont {Chmiela}\ \emph {et~al.}(2018)\citenamefont
  {Chmiela}, \citenamefont {Sauceda}, \citenamefont {M{\"u}ller},\ and\
  \citenamefont {Tkatchenko}}]{Tkatch2018}%
  \BibitemOpen
  \bibfield  {author} {\bibinfo {author} {\bibfnamefont {S.}~\bibnamefont
  {Chmiela}}, \bibinfo {author} {\bibfnamefont {H.~E.}\ \bibnamefont
  {Sauceda}}, \bibinfo {author} {\bibfnamefont {K.-R.}\ \bibnamefont
  {M{\"u}ller}}, \ and\ \bibinfo {author} {\bibfnamefont {A.}~\bibnamefont
  {Tkatchenko}},\ }\href {\doibase 10.1038/s41467-018-06169-2} {\bibfield
  {journal} {\bibinfo  {journal} {Nat. Commun.}\ }\textbf {\bibinfo {volume}
  {9}},\ \bibinfo {pages} {3887} (\bibinfo {year} {2018})}\BibitemShut
  {NoStop}%
\bibitem [{\citenamefont {Sauceda}\ \emph
  {et~al.}(2019{\natexlab{b}})\citenamefont {Sauceda}, \citenamefont {Chmiela},
  \citenamefont {Poltavsky}, \citenamefont {Müller},\ and\ \citenamefont
  {Tkatchenko}}]{Tkatch19}%
  \BibitemOpen
  \bibfield  {author} {\bibinfo {author} {\bibfnamefont {H.~E.}\ \bibnamefont
  {Sauceda}}, \bibinfo {author} {\bibfnamefont {S.}~\bibnamefont {Chmiela}},
  \bibinfo {author} {\bibfnamefont {I.}~\bibnamefont {Poltavsky}}, \bibinfo
  {author} {\bibfnamefont {K.-R.}\ \bibnamefont {Müller}}, \ and\ \bibinfo
  {author} {\bibfnamefont {A.}~\bibnamefont {Tkatchenko}},\ }\href {\doibase
  10.1063/1.5078687} {\bibfield  {journal} {\bibinfo  {journal} {J. Chem.
  Phys.}\ }\textbf {\bibinfo {volume} {150}},\ \bibinfo {pages} {114102}
  (\bibinfo {year} {2019}{\natexlab{b}})}\BibitemShut {NoStop}%
\bibitem [{\citenamefont {Unke}\ and\ \citenamefont {Meuwly}(2019)}]{PhysNet}%
  \BibitemOpen
  \bibfield  {author} {\bibinfo {author} {\bibfnamefont {O.~T.}\ \bibnamefont
  {Unke}}\ and\ \bibinfo {author} {\bibfnamefont {M.}~\bibnamefont {Meuwly}},\
  }\href {\doibase 10.1021/acs.jctc.9b00181} {\bibfield  {journal} {\bibinfo
  {journal} {J. Chem. Theory Comput.}\ }\textbf {\bibinfo {volume} {15}},\
  \bibinfo {pages} {3678} (\bibinfo {year} {2019})}\BibitemShut {NoStop}%
\bibitem [{\citenamefont {Dral}\ \emph {et~al.}(2017)\citenamefont {Dral},
  \citenamefont {Owens}, \citenamefont {Yurchenko},\ and\ \citenamefont
  {Thiel}}]{KREG}%
  \BibitemOpen
  \bibfield  {author} {\bibinfo {author} {\bibfnamefont {P.~O.}\ \bibnamefont
  {Dral}}, \bibinfo {author} {\bibfnamefont {A.}~\bibnamefont {Owens}},
  \bibinfo {author} {\bibfnamefont {S.~N.}\ \bibnamefont {Yurchenko}}, \ and\
  \bibinfo {author} {\bibfnamefont {W.}~\bibnamefont {Thiel}},\ }\href
  {\doibase 10.1063/1.4989536} {\bibfield  {journal} {\bibinfo  {journal} {J.
  Chem. Phys.}\ }\textbf {\bibinfo {volume} {146}},\ \bibinfo {pages} {244108}
  (\bibinfo {year} {2017})}\BibitemShut {NoStop}%
\bibitem [{\citenamefont {Dral}(2019)}]{pKREG}%
  \BibitemOpen
  \bibfield  {author} {\bibinfo {author} {\bibfnamefont {P.~O.}\ \bibnamefont
  {Dral}},\ }\href {\doibase https://doi.org/10.1002/jcc.26004} {\bibfield
  {journal} {\bibinfo  {journal} {J. Comput. Chem}\ }\textbf {\bibinfo {volume}
  {40}},\ \bibinfo {pages} {2339} (\bibinfo {year} {2019})}\BibitemShut
  {NoStop}%
\bibitem [{\citenamefont {Kovács}\ \emph {et~al.}(2021)\citenamefont
  {Kovács}, \citenamefont {Oord}, \citenamefont {Kucera}, \citenamefont
  {Allen}, \citenamefont {Cole}, \citenamefont {Ortner},\ and\ \citenamefont
  {Csányi}}]{ACEPIP_21}%
  \BibitemOpen
  \bibfield  {author} {\bibinfo {author} {\bibfnamefont {D.~P.}\ \bibnamefont
  {Kovács}}, \bibinfo {author} {\bibfnamefont {C.~v.~d.}\ \bibnamefont
  {Oord}}, \bibinfo {author} {\bibfnamefont {J.}~\bibnamefont {Kucera}},
  \bibinfo {author} {\bibfnamefont {A.~E.~A.}\ \bibnamefont {Allen}}, \bibinfo
  {author} {\bibfnamefont {D.~J.}\ \bibnamefont {Cole}}, \bibinfo {author}
  {\bibfnamefont {C.}~\bibnamefont {Ortner}}, \ and\ \bibinfo {author}
  {\bibfnamefont {G.}~\bibnamefont {Csányi}},\ }\href {\doibase
  10.1021/acs.jctc.1c00647} {\bibfield  {journal} {\bibinfo  {journal} {J.
  Comput. Theory Chem.}\ }\textbf {\bibinfo {volume} {xx}},\ \bibinfo {pages}
  {xxx} (\bibinfo {year} {2021})},\ \bibinfo {note} {pMID:
  34735161}\BibitemShut {NoStop}%
\bibitem [{\citenamefont {Rupp}\ \emph {et~al.}(2012)\citenamefont {Rupp},
  \citenamefont {Tkatchenko}, \citenamefont {M\"uller},\ and\ \citenamefont
  {von Lilienfeld}}]{cmatrix}%
  \BibitemOpen
  \bibfield  {author} {\bibinfo {author} {\bibfnamefont {M.}~\bibnamefont
  {Rupp}}, \bibinfo {author} {\bibfnamefont {A.}~\bibnamefont {Tkatchenko}},
  \bibinfo {author} {\bibfnamefont {K.-R.}\ \bibnamefont {M\"uller}}, \ and\
  \bibinfo {author} {\bibfnamefont {O.~A.}\ \bibnamefont {von Lilienfeld}},\
  }\href {\doibase 10.1103/PhysRevLett.108.058301} {\bibfield  {journal}
  {\bibinfo  {journal} {Phys. Rev. Lett.}\ }\textbf {\bibinfo {volume} {108}},\
  \bibinfo {pages} {058301} (\bibinfo {year} {2012})}\BibitemShut {NoStop}%
\bibitem [{\citenamefont {De}\ \emph {et~al.}(2016)\citenamefont {De},
  \citenamefont {Bart\'ok}, \citenamefont {Cs\'anyi},\ and\ \citenamefont
  {Ceriotti}}]{soap16}%
  \BibitemOpen
  \bibfield  {author} {\bibinfo {author} {\bibfnamefont {S.}~\bibnamefont
  {De}}, \bibinfo {author} {\bibfnamefont {A.~P.}\ \bibnamefont {Bart\'ok}},
  \bibinfo {author} {\bibfnamefont {G.}~\bibnamefont {Cs\'anyi}}, \ and\
  \bibinfo {author} {\bibfnamefont {M.}~\bibnamefont {Ceriotti}},\ }\href
  {\doibase 10.1039/c6cp00415f} {\bibfield  {journal} {\bibinfo  {journal}
  {Phys. Chem. Chem. Phys.}\ }\textbf {\bibinfo {volume} {18}},\ \bibinfo
  {pages} {13754} (\bibinfo {year} {2016})}\BibitemShut {NoStop}%
\bibitem [{\citenamefont {Bart\'ok}, \citenamefont {Kondor},\ and\
  \citenamefont {Cs\'anyi}(2013)}]{GP-2013}%
  \BibitemOpen
  \bibfield  {author} {\bibinfo {author} {\bibfnamefont {A.~P.}\ \bibnamefont
  {Bart\'ok}}, \bibinfo {author} {\bibfnamefont {R.}~\bibnamefont {Kondor}}, \
  and\ \bibinfo {author} {\bibfnamefont {G.}~\bibnamefont {Cs\'anyi}},\ }\href
  {\doibase 10.1103/PhysRevB.87.184115} {\bibfield  {journal} {\bibinfo
  {journal} {Phys. Rev. B}\ }\textbf {\bibinfo {volume} {87}},\ \bibinfo
  {pages} {184115} (\bibinfo {year} {2013})}\BibitemShut {NoStop}%
\bibitem [{\citenamefont {Behler}(2015)}]{Behler15}%
  \BibitemOpen
  \bibfield  {author} {\bibinfo {author} {\bibfnamefont {J.}~\bibnamefont
  {Behler}},\ }\href {\doibase 10.1002/qua.24890} {\bibfield  {journal}
  {\bibinfo  {journal} {Int. J. Quantum Chem.}\ }\textbf {\bibinfo {volume}
  {115}},\ \bibinfo {pages} {1032} (\bibinfo {year} {2015})}\BibitemShut
  {NoStop}%
\bibitem [{\citenamefont {Sch\"{u}tt}\ \emph {et~al.}(2018)\citenamefont
  {Sch\"{u}tt}, \citenamefont {Sauceda}, \citenamefont {Kindermans},
  \citenamefont {Tkatchenko},\ and\ \citenamefont {M\"{u}ller}}]{SchNet}%
  \BibitemOpen
  \bibfield  {author} {\bibinfo {author} {\bibfnamefont {K.~T.}\ \bibnamefont
  {Sch\"{u}tt}}, \bibinfo {author} {\bibfnamefont {H.~E.}\ \bibnamefont
  {Sauceda}}, \bibinfo {author} {\bibfnamefont {P.-J.}\ \bibnamefont
  {Kindermans}}, \bibinfo {author} {\bibfnamefont {A.}~\bibnamefont
  {Tkatchenko}}, \ and\ \bibinfo {author} {\bibfnamefont {K.-R.}\ \bibnamefont
  {M\"{u}ller}},\ }\href {\doibase 10.1063/1.5019779} {\bibfield  {journal}
  {\bibinfo  {journal} {J. Chem. Phys.}\ }\textbf {\bibinfo {volume} {148}},\
  \bibinfo {pages} {241722} (\bibinfo {year} {2018})}\BibitemShut {NoStop}%
\bibitem [{\citenamefont {Chen}\ \emph {et~al.}(2019)\citenamefont {Chen},
  \citenamefont {Ye}, \citenamefont {Zuo}, \citenamefont {Zheng},\ and\
  \citenamefont {Ong}}]{2019Chen}%
  \BibitemOpen
  \bibfield  {author} {\bibinfo {author} {\bibfnamefont {C.}~\bibnamefont
  {Chen}}, \bibinfo {author} {\bibfnamefont {W.}~\bibnamefont {Ye}}, \bibinfo
  {author} {\bibfnamefont {Y.}~\bibnamefont {Zuo}}, \bibinfo {author}
  {\bibfnamefont {C.}~\bibnamefont {Zheng}}, \ and\ \bibinfo {author}
  {\bibfnamefont {S.~P.}\ \bibnamefont {Ong}},\ }\href {\doibase
  10.1021/acs.chemmater.9b01294} {\bibfield  {journal} {\bibinfo  {journal}
  {Chem. Mater.}\ }\textbf {\bibinfo {volume} {31}},\ \bibinfo {pages}
  {3564–3572} (\bibinfo {year} {2019})}\BibitemShut {NoStop}%
\bibitem [{\citenamefont {Houston}\ \emph {et~al.}(2020)\citenamefont
  {Houston}, \citenamefont {Conte}, \citenamefont {Qu},\ and\ \citenamefont
  {Bowman}}]{tropolone20}%
  \BibitemOpen
  \bibfield  {author} {\bibinfo {author} {\bibfnamefont {P.~L.}\ \bibnamefont
  {Houston}}, \bibinfo {author} {\bibfnamefont {R.}~\bibnamefont {Conte}},
  \bibinfo {author} {\bibfnamefont {C.}~\bibnamefont {Qu}}, \ and\ \bibinfo
  {author} {\bibfnamefont {J.~M.}\ \bibnamefont {Bowman}},\ }\href@noop {}
  {\bibfield  {journal} {\bibinfo  {journal} {J. Chem. Phys.}\ }\textbf
  {\bibinfo {volume} {153}},\ \bibinfo {pages} {024107} (\bibinfo {year}
  {2020})}\BibitemShut {NoStop}%
\bibitem [{\citenamefont {Stöhr}, \citenamefont {Medrano~Sandonas},\ and\
  \citenamefont {Tkatchenko}(2020)}]{glycine20}%
  \BibitemOpen
  \bibfield  {author} {\bibinfo {author} {\bibfnamefont {M.}~\bibnamefont
  {Stöhr}}, \bibinfo {author} {\bibfnamefont {L.}~\bibnamefont
  {Medrano~Sandonas}}, \ and\ \bibinfo {author} {\bibfnamefont
  {A.}~\bibnamefont {Tkatchenko}},\ }\href {\doibase
  10.1021/acs.jpclett.0c01307} {\bibfield  {journal} {\bibinfo  {journal} {J.
  Phys. Chem. Lett.}\ }\textbf {\bibinfo {volume} {11}},\ \bibinfo {pages}
  {6835} (\bibinfo {year} {2020})}\BibitemShut {NoStop}%
\bibitem [{\citenamefont {Conte}\ \emph
  {et~al.}(2020{\natexlab{b}})\citenamefont {Conte}, \citenamefont {Houston},
  \citenamefont {Qu}, \citenamefont {Li},\ and\ \citenamefont
  {Bowman}}]{conte_glycine20}%
  \BibitemOpen
  \bibfield  {author} {\bibinfo {author} {\bibfnamefont {R.}~\bibnamefont
  {Conte}}, \bibinfo {author} {\bibfnamefont {P.~L.}\ \bibnamefont {Houston}},
  \bibinfo {author} {\bibfnamefont {C.}~\bibnamefont {Qu}}, \bibinfo {author}
  {\bibfnamefont {J.}~\bibnamefont {Li}}, \ and\ \bibinfo {author}
  {\bibfnamefont {J.~M.}\ \bibnamefont {Bowman}},\ }\href {\doibase
  10.1063/5.0037175} {\bibfield  {journal} {\bibinfo  {journal} {J. Chem.
  Phys.}\ }\textbf {\bibinfo {volume} {153}},\ \bibinfo {pages} {244301}
  (\bibinfo {year} {2020}{\natexlab{b}})}\BibitemShut {NoStop}%
\bibitem [{\citenamefont {Nandi}, \citenamefont {Qu},\ and\ \citenamefont
  {Bowman}(2019)}]{NandiQuBowman2019}%
  \BibitemOpen
  \bibfield  {author} {\bibinfo {author} {\bibfnamefont {A.}~\bibnamefont
  {Nandi}}, \bibinfo {author} {\bibfnamefont {C.}~\bibnamefont {Qu}}, \ and\
  \bibinfo {author} {\bibfnamefont {J.~M.}\ \bibnamefont {Bowman}},\
  }\href@noop {} {\bibfield  {journal} {\bibinfo  {journal} {J. Chem. Theor.
  Comp.}\ }\textbf {\bibinfo {volume} {15}},\ \bibinfo {pages} {2826} (\bibinfo
  {year} {2019})}\BibitemShut {NoStop}%
\bibitem [{\citenamefont {Nandi}\ \emph
  {et~al.}(2021{\natexlab{a}})\citenamefont {Nandi}, \citenamefont {Qu},
  \citenamefont {Houston}, \citenamefont {Conte}, \citenamefont {Yu},\ and\
  \citenamefont {Bowman}}]{4body}%
  \BibitemOpen
  \bibfield  {author} {\bibinfo {author} {\bibfnamefont {A.}~\bibnamefont
  {Nandi}}, \bibinfo {author} {\bibfnamefont {C.}~\bibnamefont {Qu}}, \bibinfo
  {author} {\bibfnamefont {P.~L.}\ \bibnamefont {Houston}}, \bibinfo {author}
  {\bibfnamefont {R.}~\bibnamefont {Conte}}, \bibinfo {author} {\bibfnamefont
  {Q.}~\bibnamefont {Yu}}, \ and\ \bibinfo {author} {\bibfnamefont {J.~M.}\
  \bibnamefont {Bowman}},\ }\href {\doibase 10.1021/acs.jpclett.1c03152}
  {\bibfield  {journal} {\bibinfo  {journal} {J. Phys. Chem. Lett.}\ }\textbf
  {\bibinfo {volume} {12}},\ \bibinfo {pages} {10318} (\bibinfo {year}
  {2021}{\natexlab{a}})}\BibitemShut {NoStop}%
\bibitem [{\citenamefont {Braams}\ and\ \citenamefont
  {Bowman}(2009)}]{Braams2009}%
  \BibitemOpen
  \bibfield  {author} {\bibinfo {author} {\bibfnamefont {B.~J.}\ \bibnamefont
  {Braams}}\ and\ \bibinfo {author} {\bibfnamefont {J.~M.}\ \bibnamefont
  {Bowman}},\ }\href {\doibase 10.1080/01442350903234923} {\bibfield  {journal}
  {\bibinfo  {journal} {Int. Rev. Phys. Chem.}\ }\textbf {\bibinfo {volume}
  {28}},\ \bibinfo {pages} {577} (\bibinfo {year} {2009})}\BibitemShut
  {NoStop}%
\bibitem [{Xie(2019)}]{Xie-nma}%
  \BibitemOpen
  \href {https://www.mcs.anl.gov/research/projects/msa/} {\enquote {\bibinfo
  {title} {Original msa software},}\ }\bibinfo {howpublished}
  {\url{https://www.mcs.anl.gov/research/projects/msa/}} (\bibinfo {year}
  {2019}),\ \bibinfo {note} {accessed: 2019-12-20}\BibitemShut {NoStop}%
\bibitem [{msa(2019)}]{msachen}%
  \BibitemOpen
  \href@noop {} {\enquote {\bibinfo {title} {Msa software with gradients},}\
  }\bibinfo {howpublished} {\url{https://github.com/szquchen/MSA-2.0}}
  (\bibinfo {year} {2019}),\ \bibinfo {note} {accessed: 2019-01-20}\BibitemShut
  {NoStop}%
\bibitem [{\citenamefont {Conte}, \citenamefont {Houston},\ and\ \citenamefont
  {Bowman}(2014)}]{purified14}%
  \BibitemOpen
  \bibfield  {author} {\bibinfo {author} {\bibfnamefont {R.}~\bibnamefont
  {Conte}}, \bibinfo {author} {\bibfnamefont {P.~L.}\ \bibnamefont {Houston}},
  \ and\ \bibinfo {author} {\bibfnamefont {J.~M.}\ \bibnamefont {Bowman}},\
  }\href@noop {} {\bibfield  {journal} {\bibinfo  {journal} {J. Chem. Phys.}\
  }\textbf {\bibinfo {volume} {140}},\ \bibinfo {pages} {151101} (\bibinfo
  {year} {2014})}\BibitemShut {NoStop}%
\bibitem [{\citenamefont {Conte}, \citenamefont {Qu},\ and\ \citenamefont
  {Bowman}(2015)}]{ConteQuBowman2015}%
  \BibitemOpen
  \bibfield  {author} {\bibinfo {author} {\bibfnamefont {R.}~\bibnamefont
  {Conte}}, \bibinfo {author} {\bibfnamefont {C.}~\bibnamefont {Qu}}, \ and\
  \bibinfo {author} {\bibfnamefont {J.~M.}\ \bibnamefont {Bowman}},\
  }\href@noop {} {\bibfield  {journal} {\bibinfo  {journal} {J. Chem. Theory
  Comput.}\ }\textbf {\bibinfo {volume} {11}},\ \bibinfo {pages} {1631}
  (\bibinfo {year} {2015})}\BibitemShut {NoStop}%
\bibitem [{\citenamefont {Homayoon}\ \emph {et~al.}(2015)\citenamefont
  {Homayoon}, \citenamefont {Conte}, \citenamefont {Qu},\ and\ \citenamefont
  {Bowman}}]{purified15c}%
  \BibitemOpen
  \bibfield  {author} {\bibinfo {author} {\bibfnamefont {Z.}~\bibnamefont
  {Homayoon}}, \bibinfo {author} {\bibfnamefont {R.}~\bibnamefont {Conte}},
  \bibinfo {author} {\bibfnamefont {C.}~\bibnamefont {Qu}}, \ and\ \bibinfo
  {author} {\bibfnamefont {J.~M.}\ \bibnamefont {Bowman}},\ }\href {\doibase
  10.1063/1.4929338} {\bibfield  {journal} {\bibinfo  {journal} {J. Chem.
  Phys.}\ }\textbf {\bibinfo {volume} {143}},\ \bibinfo {pages} {084302}
  (\bibinfo {year} {2015})}\BibitemShut {NoStop}%
\bibitem [{\citenamefont {Qu}\ \emph {et~al.}(2015)\citenamefont {Qu},
  \citenamefont {Conte}, \citenamefont {Houston},\ and\ \citenamefont
  {Bowman}}]{QuConteHoustonBowman2015}%
  \BibitemOpen
  \bibfield  {author} {\bibinfo {author} {\bibfnamefont {C.}~\bibnamefont
  {Qu}}, \bibinfo {author} {\bibfnamefont {R.}~\bibnamefont {Conte}}, \bibinfo
  {author} {\bibfnamefont {P.~L.}\ \bibnamefont {Houston}}, \ and\ \bibinfo
  {author} {\bibfnamefont {J.~M.}\ \bibnamefont {Bowman}},\ }\href {\doibase
  10.1039/c4cp05913a} {\bibfield  {journal} {\bibinfo  {journal} {Phys. Chem.
  Chem. Phys.}\ }\textbf {\bibinfo {volume} {17}},\ \bibinfo {pages} {8172}
  (\bibinfo {year} {2015})}\BibitemShut {NoStop}%
\bibitem [{\citenamefont {Paukku}\ \emph {et~al.}(2013)\citenamefont {Paukku},
  \citenamefont {Yang}, \citenamefont {Varga},\ and\ \citenamefont
  {Truhlar}}]{purified13}%
  \BibitemOpen
  \bibfield  {author} {\bibinfo {author} {\bibfnamefont {Y.}~\bibnamefont
  {Paukku}}, \bibinfo {author} {\bibfnamefont {K.~R.}\ \bibnamefont {Yang}},
  \bibinfo {author} {\bibfnamefont {Z.}~\bibnamefont {Varga}}, \ and\ \bibinfo
  {author} {\bibfnamefont {D.~G.}\ \bibnamefont {Truhlar}},\ }\href {\doibase
  10.1063/1.4811653} {\bibfield  {journal} {\bibinfo  {journal} {J. Chem.
  Phys.}\ }\textbf {\bibinfo {volume} {139}},\ \bibinfo {pages} {044309}
  (\bibinfo {year} {2013})}\BibitemShut {NoStop}%
\bibitem [{\citenamefont {Qu}\ and\ \citenamefont
  {Bowman}(2019)}]{QuBowman2019}%
  \BibitemOpen
  \bibfield  {author} {\bibinfo {author} {\bibfnamefont {C.}~\bibnamefont
  {Qu}}\ and\ \bibinfo {author} {\bibfnamefont {J.~M.}\ \bibnamefont
  {Bowman}},\ }\href@noop {} {\bibfield  {journal} {\bibinfo  {journal} {J.
  Chem. Phys.}\ }\textbf {\bibinfo {volume} {150}},\ \bibinfo {pages} {141101}
  (\bibinfo {year} {2019})}\BibitemShut {NoStop}%
\bibitem [{\citenamefont
  {Wolfram\hspace{.1cm}Research\hspace{.1cm}Inc.}(2019)}]{Mathematica}%
  \BibitemOpen
  \bibfield  {author} {\bibinfo {author} {\bibnamefont
  {Wolfram\hspace{.1cm}Research\hspace{.1cm}Inc.}},\ }\href@noop {} {\enquote
  {\bibinfo {title} {Mathematica, {V}ersion 12.0},}\ } (\bibinfo {year}
  {2019}),\ \bibinfo {note} {champaign, IL, 2019}\BibitemShut {NoStop}%
\bibitem [{\citenamefont {Baydin}\ and\ \citenamefont
  {Pearlmutter}(2014)}]{BaydinPearlmutter2014}%
  \BibitemOpen
  \bibfield  {author} {\bibinfo {author} {\bibfnamefont {A.~G.}\ \bibnamefont
  {Baydin}}\ and\ \bibinfo {author} {\bibfnamefont {B.~A.}\ \bibnamefont
  {Pearlmutter}},\ }\href@noop {} {\bibfield  {journal} {\bibinfo  {journal}
  {JMLR: Workshop and Conference Proceedings, ICML 2014 AutoML Workshop}\ ,\
  \bibinfo {pages} {1}} (\bibinfo {year} {2014})}\BibitemShut {NoStop}%
\bibitem [{\citenamefont {Baydin}\ \emph {et~al.}(2017)\citenamefont {Baydin},
  \citenamefont {Pearlmutter}, \citenamefont {Radul},\ and\ \citenamefont
  {Siskind}}]{Baydin2018}%
  \BibitemOpen
  \bibfield  {author} {\bibinfo {author} {\bibfnamefont {A.~G.}\ \bibnamefont
  {Baydin}}, \bibinfo {author} {\bibfnamefont {B.~A.}\ \bibnamefont
  {Pearlmutter}}, \bibinfo {author} {\bibfnamefont {A.~A.}\ \bibnamefont
  {Radul}}, \ and\ \bibinfo {author} {\bibfnamefont {J.~M.}\ \bibnamefont
  {Siskind}},\ }\href@noop {} {\bibfield  {journal} {\bibinfo  {journal} {The
  Journal of Machine Learning Research}\ }\textbf {\bibinfo {volume} {18}},\
  \bibinfo {pages} {5595–5637} (\bibinfo {year} {2017})}\BibitemShut
  {NoStop}%
\bibitem [{\citenamefont {Abbott}\ \emph {et~al.}(2021)\citenamefont {Abbott},
  \citenamefont {Abbott}, \citenamefont {Turney},\ and\ \citenamefont
  {Schaefer}}]{autodiffSchaefer}%
  \BibitemOpen
  \bibfield  {author} {\bibinfo {author} {\bibfnamefont {A.~S.}\ \bibnamefont
  {Abbott}}, \bibinfo {author} {\bibfnamefont {B.~Z.}\ \bibnamefont {Abbott}},
  \bibinfo {author} {\bibfnamefont {J.~M.}\ \bibnamefont {Turney}}, \ and\
  \bibinfo {author} {\bibfnamefont {H.~F.}\ \bibnamefont {Schaefer}},\ }\href
  {\doibase 10.1021/acs.jpclett.1c00607} {\bibfield  {journal} {\bibinfo
  {journal} {J. Phys. Chem. Lett.}\ }\textbf {\bibinfo {volume} {12}},\
  \bibinfo {pages} {3232} (\bibinfo {year} {2021})}\BibitemShut {NoStop}%
\bibitem [{\citenamefont {Griewank}\ and\ \citenamefont
  {Walther}(2008)}]{GriewankWalther2008}%
  \BibitemOpen
  \bibfield  {author} {\bibinfo {author} {\bibfnamefont {A.}~\bibnamefont
  {Griewank}}\ and\ \bibinfo {author} {\bibfnamefont {A.}~\bibnamefont
  {Walther}},\ }\href {\doibase 10.1137/1.9780898717761} {\emph {\bibinfo
  {title} {Evaluating Derivatives: Principles and Techniques of Algorithmic
  Differentiation.}}}\ (\bibinfo  {publisher} {Society for Industrial and
  Applied Mathematics},\ \bibinfo {address} {Philadelphia},\ \bibinfo {year}
  {2008})\BibitemShut {NoStop}%
\bibitem [{\citenamefont {Zheng}\ \emph {et~al.}(2011)\citenamefont {Zheng},
  \citenamefont {Yu}, \citenamefont {Papajak}, \citenamefont {Alecu},
  \citenamefont {Mielke},\ and\ \citenamefont {Truhlar}}]{ethpartit}%
  \BibitemOpen
  \bibfield  {author} {\bibinfo {author} {\bibfnamefont {J.}~\bibnamefont
  {Zheng}}, \bibinfo {author} {\bibfnamefont {T.}~\bibnamefont {Yu}}, \bibinfo
  {author} {\bibfnamefont {E.}~\bibnamefont {Papajak}}, \bibinfo {author}
  {\bibfnamefont {I.~M.}\ \bibnamefont {Alecu}}, \bibinfo {author}
  {\bibfnamefont {S.~L.}\ \bibnamefont {Mielke}}, \ and\ \bibinfo {author}
  {\bibfnamefont {D.~G.}\ \bibnamefont {Truhlar}},\ }\href {\doibase
  10.1039/C0CP02644A} {\bibfield  {journal} {\bibinfo  {journal} {Phys. Chem.
  Chem. Phys.}\ }\textbf {\bibinfo {volume} {13}},\ \bibinfo {pages} {10885}
  (\bibinfo {year} {2011})}\BibitemShut {NoStop}%
\bibitem [{\citenamefont {Werner}\ \emph {et~al.}(2015)\citenamefont {Werner},
  \citenamefont {Knowles}, \citenamefont {Knizia}, \citenamefont {Manby},\ and\
  \citenamefont {{Sch\"{u}tz}}}]{MOLPRO_brief}%
  \BibitemOpen
  \bibfield  {author} {\bibinfo {author} {\bibfnamefont {H.-J.}\ \bibnamefont
  {Werner}}, \bibinfo {author} {\bibfnamefont {P.~J.}\ \bibnamefont {Knowles}},
  \bibinfo {author} {\bibfnamefont {G.}~\bibnamefont {Knizia}}, \bibinfo
  {author} {\bibfnamefont {F.~R.}\ \bibnamefont {Manby}}, \ and\ \bibinfo
  {author} {\bibfnamefont {M.}~\bibnamefont {{Sch\"{u}tz}}},\ }\href@noop {}
  {\enquote {\bibinfo {title} {Molpro, version 2015.1, a package of ab initio
  programs},}\ } (\bibinfo {year} {2015}),\ \bibinfo {note} {see
  http://www.molpro.net}\BibitemShut {NoStop}%
\bibitem [{\citenamefont {Qu}\ \emph {et~al.}(2020)\citenamefont {Qu},
  \citenamefont {Conte}, \citenamefont {Houston},\ and\ \citenamefont
  {Bowman}}]{QuAcAc}%
  \BibitemOpen
  \bibfield  {author} {\bibinfo {author} {\bibfnamefont {C.}~\bibnamefont
  {Qu}}, \bibinfo {author} {\bibfnamefont {R.}~\bibnamefont {Conte}}, \bibinfo
  {author} {\bibfnamefont {P.~L.}\ \bibnamefont {Houston}}, \ and\ \bibinfo
  {author} {\bibfnamefont {J.~M.}\ \bibnamefont {Bowman}},\ }\href {\doibase
  10.1039/D0CP04221H} {\bibfield  {journal} {\bibinfo  {journal} {Phys. Chem.
  Chem. Phys.}\ ,\ } (\bibinfo {year} {2020})}\BibitemShut {NoStop}%
\bibitem [{\citenamefont {K{\"a}ser}, \citenamefont {Unke},\ and\ \citenamefont
  {Meuwly}(2020)}]{meuwly2020}%
  \BibitemOpen
  \bibfield  {author} {\bibinfo {author} {\bibfnamefont {S.}~\bibnamefont
  {K{\"a}ser}}, \bibinfo {author} {\bibfnamefont {O.}~\bibnamefont {Unke}}, \
  and\ \bibinfo {author} {\bibfnamefont {M.}~\bibnamefont {Meuwly}},\
  }\href@noop {} {\bibfield  {journal} {\bibinfo  {journal} {New Journal of
  Physics}\ }\textbf {\bibinfo {volume} {22}},\ \bibinfo {pages} {055002}
  (\bibinfo {year} {2020})}\BibitemShut {NoStop}%
\bibitem [{\citenamefont {Fortenberry}\ \emph {et~al.}(2015)\citenamefont
  {Fortenberry}, \citenamefont {Yu}, \citenamefont {Mancini}, \citenamefont
  {Bowman}, \citenamefont {Lee}, \citenamefont {Crawford}, \citenamefont
  {Klemperer},\ and\ \citenamefont {Francisco}}]{OCHCO15}%
  \BibitemOpen
  \bibfield  {author} {\bibinfo {author} {\bibfnamefont {R.~C.}\ \bibnamefont
  {Fortenberry}}, \bibinfo {author} {\bibfnamefont {Q.}~\bibnamefont {Yu}},
  \bibinfo {author} {\bibfnamefont {J.~S.}\ \bibnamefont {Mancini}}, \bibinfo
  {author} {\bibfnamefont {J.~M.}\ \bibnamefont {Bowman}}, \bibinfo {author}
  {\bibfnamefont {T.~J.}\ \bibnamefont {Lee}}, \bibinfo {author} {\bibfnamefont
  {T.~D.}\ \bibnamefont {Crawford}}, \bibinfo {author} {\bibfnamefont {W.~F.}\
  \bibnamefont {Klemperer}}, \ and\ \bibinfo {author} {\bibfnamefont {J.~S.}\
  \bibnamefont {Francisco}},\ }\href {\doibase 10.1063/1.4929345} {\bibfield
  {journal} {\bibinfo  {journal} {J. Chem. Phys.}\ }\textbf {\bibinfo {volume}
  {143}},\ \bibinfo {pages} {071102} (\bibinfo {year} {2015})}\BibitemShut
  {NoStop}%
\bibitem [{\citenamefont {Brorsen}(2019)}]{brorsen19}%
  \BibitemOpen
  \bibfield  {author} {\bibinfo {author} {\bibfnamefont {K.~R.}\ \bibnamefont
  {Brorsen}},\ }\href {\doibase 10.1063/1.5093908} {\bibfield  {journal}
  {\bibinfo  {journal} {J. Chem. Phys.}\ }\textbf {\bibinfo {volume} {150}},\
  \bibinfo {pages} {204104} (\bibinfo {year} {2019})}\BibitemShut {NoStop}%
\bibitem [{\citenamefont {Xie}\ and\ \citenamefont {Bowman}(2010)}]{Xie10}%
  \BibitemOpen
  \bibfield  {author} {\bibinfo {author} {\bibfnamefont {Z.}~\bibnamefont
  {Xie}}\ and\ \bibinfo {author} {\bibfnamefont {J.~M.}\ \bibnamefont
  {Bowman}},\ }\href {\doibase 10.1021/ct9004917} {\bibfield  {journal}
  {\bibinfo  {journal} {J. Chem. Theory Comput.}\ }\textbf {\bibinfo {volume}
  {6}},\ \bibinfo {pages} {26} (\bibinfo {year} {2010})}\BibitemShut {NoStop}%
\bibitem [{\citenamefont {Nandi}\ \emph
  {et~al.}(2021{\natexlab{b}})\citenamefont {Nandi}, \citenamefont {Qu},
  \citenamefont {Houston}, \citenamefont {Conte},\ and\ \citenamefont
  {Bowman}}]{nandidelta_21}%
  \BibitemOpen
  \bibfield  {author} {\bibinfo {author} {\bibfnamefont {A.}~\bibnamefont
  {Nandi}}, \bibinfo {author} {\bibfnamefont {C.}~\bibnamefont {Qu}}, \bibinfo
  {author} {\bibfnamefont {P.~L.}\ \bibnamefont {Houston}}, \bibinfo {author}
  {\bibfnamefont {R.}~\bibnamefont {Conte}}, \ and\ \bibinfo {author}
  {\bibfnamefont {J.~M.}\ \bibnamefont {Bowman}},\ }\href {\doibase
  10.1063/5.0038301} {\bibfield  {journal} {\bibinfo  {journal} {J. Chem.
  Phys.}\ }\textbf {\bibinfo {volume} {154}},\ \bibinfo {pages} {051102}
  (\bibinfo {year} {2021}{\natexlab{b}})}\BibitemShut {NoStop}%
\bibitem [{\citenamefont {Qu}\ \emph {et~al.}(2021)\citenamefont {Qu},
  \citenamefont {Houston}, \citenamefont {Conte}, \citenamefont {Nandi},\ and\
  \citenamefont {Bowman}}]{QudeltaJPCL_21}%
  \BibitemOpen
  \bibfield  {author} {\bibinfo {author} {\bibfnamefont {C.}~\bibnamefont
  {Qu}}, \bibinfo {author} {\bibfnamefont {P.~L.}\ \bibnamefont {Houston}},
  \bibinfo {author} {\bibfnamefont {R.}~\bibnamefont {Conte}}, \bibinfo
  {author} {\bibfnamefont {A.}~\bibnamefont {Nandi}}, \ and\ \bibinfo {author}
  {\bibfnamefont {J.~M.}\ \bibnamefont {Bowman}},\ }\href {\doibase
  10.1021/acs.jpclett.1c01142} {\bibfield  {journal} {\bibinfo  {journal} {J.
  Phys. Chem. Lett.}\ }\textbf {\bibinfo {volume} {12}},\ \bibinfo {pages}
  {4902} (\bibinfo {year} {2021})}\BibitemShut {NoStop}%
\end{thebibliography}%
\end{document}


\title{Supplementary Material
\date{\today}
\newline
\newline
Permutationally invariant polynomial regression for fitting energies and gradients, using reverse differentiation, achieves orders of magnitude speed-up with high precision compared to other machine learning methods}
\author{Paul L. Houston}
\email{plh2@cornell.edu}
\affiliation{Department of Chemistry and Chemical Biology, Cornell University, Ithaca, New York
14853, U.S.A. and Department of Chemistry and Biochemistry, Georgia Institute of
Technology, Atlanta, Georgia 30332, U.S.A}
\author{Chen Qu}
\affiliation{Department of Chemistry \& Biochemistry, University of Maryland, College Park, Maryland 20742, U.S.A.}
\author{Apurba Nandi}
\email{apurba.nandi@emory.edu}
\affiliation{Department of Chemistry and Cherry L. Emerson Center for Scientific Computation, Emory University, Atlanta, Georgia 30322, U.S.A.}
\author{Riccardo Conte}
\email{riccardo.conte1@unimi.it}
\affiliation{Dipartimento di Chimica, Universit\`{a} Degli Studi di Milano, via Golgi 19, 20133 Milano, Italy}
\author{Qi Yu}
\email{q.yu@yale.edu}
\affiliation{Department of Chemistry, Yale University, New Haven, Connecticut, U.S.A.}
\author{Joel M. Bowman}
\email{jmbowma@emory.edu}
\affiliation{Department of Chemistry and Cherry L. Emerson Center for Scientific Computation, Emory University, Atlanta, Georgia 30322, U.S.A.}


\begin{abstract}
\end{abstract}
\maketitle
\section*{Introduction}
This supplementary material contains examples of the backward differentiation for PIP bases for a diatomic and triatomic molecule, Mathematica code,\cite{Mathematica}, a brief description of Mathematica programs uses, and training and testing RMS errors for energies and gradients, as well as timing results, for the MD17 ethanol potential. Results validating the new PES for ethanol are also included.

\section*{Reverse Derivative Examples}
We provide here two very simple worked examples to illustrate the principles of the reverse derivative technique for PIPs.  The first is the case of a homonuclear diatomic molecule, while the second is the case of a single water molecule.

Before continuing, an explanation of notation might be useful.  In working the examples, we will often talk about partial derivatives, whereas the examples provided in the two tables below show Fortran code in which the derivatives are apparently normal ones.  Of course, the end result, how the potential varies with changes in each of the Cartesian coordinates must be expressed as partial derivatives.  However, Fortran does not distinguish partial from normal derivatives, so the tables might appear at first to be at odds with the explanations.  In most cases, the derivatives are actually partial ones; usually it is clear from the context.

\subsection*{A homonuclear diatomic molecule}

Consider the case of a homonuclear diatomic molecule with maximum polynomial order of 3. The permutational symmetry is 2, and the $MSA$ code\cite{Xie-nma,msachen,Xie10,persp9} for the energy is listed in the first column of Table~SM-I. To calculate the energy for a particular geometry, one moves forward (upward) in the column by calculating first the value of the transformed inter nuclear distance, $\tau(1)$, then the $m$ values, then the $p$ values and finally the potential $V=\bm{c}\cdot\bm{p}$.

\begin{table}[htbp!]
\caption*{Table SM-I. Energy and both Forward and Reverse Automatic Differentiation for a Homonuclear Diatomic Molecule }
\centering
\begin{scriptsize}
\begin{tabular*}{\textwidth}{lll}
\hline
\hline\noalign{\smallskip}

Forward (up) & \hspace{.5 cm}Forward (up) &\hspace{1cm} Reverse (down)\\
$V=\bm{c}\cdot\bm{p}$ &
\hspace{.5 cm} $\partial{V} = \bm{c}\cdot\partial{\bm{p}}$ &
\hspace{1cm} ${\partial{V}} = \bm{c}\cdot\partial{\bm{p}}$\\
\noalign{\smallskip}\hline\noalign{\smallskip}
$p(3)=p(1)*p(2)$&
\hspace{.5 cm} $dp(3) = dp(1)*p(2) + p(1)*dp(2)$ &
\hspace{1cm} $a(6)=(dV/dp(3)) = c(3)$\\
$p(2)=p(1)*p(1)$  &
\hspace{.5 cm} $dp(2) = dp(1)*p(1) + p(1)*dp(1)$ &
\hspace{1cm} $a(5)=(dV/dp(2)) + a(6)*(dp(3)/dp(2))$\\
 &
 \hspace{.5 cm} &
 \hspace{1cm}\hspace{1cm}$= c(2) + a(6)*p(1)$\\
$p(1)=m(1)$ &
\hspace{.5 cm} $dp(1) = dm(1)$ &
\hspace{1cm}$a(4)=(dV/dp(1)) + a(5)*(dp(2)/dp(1))$\\  
&
\hspace{.5 cm} & 
\hspace{1.0cm}\hspace{1cm} $+ a(6)*(dp(3)/dp(1)$\\
 &
 \hspace{.5 cm} & 
 \hspace{1cm}\hspace{1cm}$ =c(1) + a(5)*2*p(1) +a(6)p(2)$\\
$p(0)=m(0)$ &
\hspace{.5 cm} $dp(0) = dm(0) = 0$ &
\hspace{1cm}$a(3)=(dV/dp(0)) + a(4)*(dp(1)/dp(0))$\\
 &
 \hspace{.5 cm} &
 \hspace{1.0cm}\hspace{1cm}$ + a(5)*(dp(2)/dp(0)) + a(6)*(dp(3)/dp(0))$\\
 &
 \hspace{.5 cm} &
 \hspace{1cm} \hspace{1cm}$= c(0)$\\
$m(1)=\tau(1)$ &
\hspace{.5 cm} $dm(1) = d\tau(1) = -(m(1)/\lambda)*\frac{\partial{r(1)}}{\partial{\alpha_n}}d\alpha_n$ & \hspace{1cm}$a(2)=a(3)*(dp(0)/dm(1)) + a(4)*(dp(1)/dm(1))$\\
 &\hspace{.5 cm} &
 \hspace{1cm} \hspace{1.cm} $ + $…$ = a(4)$\\
$m(0) = 1$ &
\hspace{.5 cm} $dm(0) = 0$ &
\hspace{1cm} $a(1)=a(2)*(dm(1)/dm(0)) +a(3)*(dp(0)/dm(0))$\\
 &
 \hspace{.5 cm} &
 \hspace{1cm}\hspace{1cm}$ + $…$=$ $ 0 + 1 = a(3)$\\

\noalign{\smallskip}\hline
\hline\noalign{\smallskip}
\end{tabular*}
\end{scriptsize}
\end{table}

The derivative steps for the calculation of the gradient of $V$ are shown in the second column of the Table~SM-I. Moving forward again, we calculate the differential of $\tau(1)$, then the differentials of the $m$, then the differentials of the $p$, and finally the differential of the potential.  Note that the differential $d\tau(1)=dm(1)$ depends on which Cartesian coordinate $\alpha_n$  we want.  In the equation given, $dm(1) = -(m(1)/\lambda)*\frac{\partial{r(1)}}{\partial{\alpha_n}}d\alpha_n$, the first factor,$ -(m(1)/\lambda)$ comes from the derivative of transformed $m(1)=\tau(1)$ with respect to the inter nuclear distance, $r(1)$, where we have assumed a Morse transform, $\tau(1)=exp(-r(1)/\lambda)$, where $\lambda$ is a range parameter generally taken to be about 2 bohr. It is instructive in this simple case to see how the derivative of the potential depends on $\tau(1)$.  One can easily verify using the definitions in the first two columns that  $\frac{\partial{V}}{\partial{\tau(1)}}=\sum_{i=1}^{3}\frac{\partial{V}}{\partial{p(i)}}\frac{\partial{p(i)}}{\partial{\tau(1)}} = \sum_{i=1}^{3}c(i)\frac{\partial{p(i)}}{\partial{\tau(1)}}=c(1) + c(2)*2*\tau(1) + c(3)*3*\tau(1)^2 $.  In order to get all six partial derivatives of $V$, we would have to evaluate each of $3N=6$ different differentials $dm(1)=d\tau(1)$, then work our way up the middle column for each choice to get the differential of $V$.

Now consider the reverse derivative method in the third column of Table SM-I (moving down). The first adjoint is $a(6)$ with conjugate variable $p(3)$, and the derivative of $V$ with respect to $p(3)$ is $c(3)$.  For the next adjoint, $a(5)$, the conjugate variable, $dp(2)$, can contribute to a change in $V$ either directly through its contribution in the dot product or indirectly through its contribution to the change in the previously calculated $a(6)$.  The contribution from the dot product is $c(2)$, whereas the potential contribution from $a(6)$ is $a(6) (dp(3)/dp(2))$. From the definition of $dp(3)$ in the second column, we find that the $(dp(3)/dp(2)) = p(1)$, so the adjoint $a(5)$ is equal to $c(2) + a(6)*p(1)$.  Continuing down the chain, the reasoning is similar.  Note that many derivatives are zero.  The important line is that for the adjoint $a(2)$ because its conjugate variable is $\tau(1)$, so its value is that of $dV/d\tau(1)$. It is instructive to see what this derivative is in terms of $\tau(1)$. We see that $a(2) = a(4) = c(1) +2a(5)p(1) +a(6)p(2)$, which can be shown to be equal to $c(1)+2*c(1)*\tau(1)+3*c(3)*\tau(1)^2$, exactly the answer we got using the Forward differentiation.  To get the gradient we want, we use $\frac{\partial{V}}{\partial{\alpha_n}}=\frac{\partial{V}}{\partial{\tau(1))}}\frac{\partial{\tau(1)}}{\partial{\alpha_n}}=\frac{\partial{V}}{\partial{\tau(1))}}\frac{\partial{\tau(1)}}{\partial{r(1)}}\frac{\partial{r(1)}}{\partial{\alpha_n}}$. From the previous paragraph, we have already seen that the rhs of the last equation is $\frac{\partial{V}}{\partial{\tau(1))}}(-m(1)/\lambda)\frac{\partial{r(1)}}{\partial{\alpha_n}}$.  The big difference between the reverse and forward methods is that in one reverse pass we get $\frac{\partial{V}}{\partial{\tau(1))}}$ and that all we need to do to get all $3N=6$ gradients is to multiply this result by $(-m(1)/\lambda)$ and by each of the six partial derivatives $\frac{\partial{r(1)}}{\partial{\alpha_n}}$.  

\subsection*{A single water molecule}

\begin{table}[ht]
\caption*{Table SM-II. Energy and both Forward and Reverse Automatic Differentiation for a Single Water Molecule }
\centering
\begin{scriptsize}
\begin{tabular*}{\textwidth}{ l l l }
\hline
\hline\noalign{\smallskip}
Forward (up) & \hspace{.5cm}Forward (up) & \hspace{.5cm}Reverse (down) \\
$V=\bm{c}\cdot\bm{p}$ & \hspace{.5cm}$\partial{V} = \bm{c}\cdot\partial{\bm{p}}$ & \hspace{.5cm}${\partial{V}} = \bm{c}\cdot\partial{\bm{p}}$\\
\noalign{\smallskip}\hline\noalign{\smallskip}
$p(12) = p(2)*p(6)$ &\hspace{.5cm} $dp(12) = dp(2)*p(6) + p(2)*dp(6)$ &\hspace{.5cm} $a(18) = c(12)$ \\
$p(11) = p(1)*p(5) $ &\hspace{.5cm} $dp(11) = dp(1)*p(5) + p(1)*dp(5) )$ &\hspace{.5cm} $a(17) = c(11)$ \\
\hspace{.5cm} $- p(8)$  &\hspace{.5cm}  \hspace{.5cm} $- dp(8)$ &\hspace{.5cm}    \\
$p(10) = p(2)*p(4)$ &\hspace{.5cm} $dp(10) = dp(2)*p(4) + p(2)*dp(4)$ &\hspace{.5cm} $a(16) = c(10)$ \\
$p(9) = p(2)*p(5)$ &\hspace{.5cm} $dp(9) = dp(2)*p(5) + p(2)*dp(5)$ &\hspace{.5cm} $a(15) = c(9)$ \\ 
$p(8) = p(3)*p(1)$  &\hspace{.5cm} $dp(8) = dp(3)*p(1) + p(3)*dp(1)$ &\hspace{.5cm} $a(14) = c(8) + a(17)*(-1)$ \\
$p(7) = p(2)*p(3)$ &\hspace{.5cm} $dp(7) = dp(2)*p(3) + p(2)*dp(3)$ &\hspace{.5cm} $a(13) = c(7)$ \\
$p(6) = p(2)*p(2)$ &\hspace{.5cm} $dp(6) = dp(2)*p(2) + p(2)*dp(2)$ &\hspace{.5cm} $a(12) = c(6) + a(18)*p(2)$\\
$p(5) = p(1)*p(1)$ &\hspace{.5cm} $dp(5) = dp(1)*p(1) + p(1)*dp(1)$ &\hspace{.5cm} $a(11) = c(5) + a(17)*p(1) + a(15)*p(2)$ \\
\hspace{.7cm}$- p(3) - p(3)$ & \hspace{1.7cm}$- dp(3) - dp(3)$ &  \\
$p(4) = p(2)*p(1)$ &\hspace{.5cm} $ dp(4) = dp(2)*p(1) +p(2)*dp(1)$ &\hspace{.5cm} $a(10) = c(4) + a(16)*p(2)$ \\
$p(3) = m(4)$ &\hspace{.5cm} $dp(3) = dm(4)$ &\hspace{.5cm} $a(9) = c(3) + a(14)*p(1) + a(13)*p(2) + $ \\
 &\hspace{.5cm} &\hspace{.5cm}\hspace{1cm} $a(11)*(-2)$ \\
$p(2) = m(3)$ &\hspace{.5cm} $dp(2) = dm(3)$ &\hspace{.5cm} $a(8) = c(2) + a(18)*p(6) + a(16)*p(4) +  $ \\
 &\hspace{.5cm} & \hspace{.5cm}\hspace{1cm} $a(15)*p(5) + a(13)*p(3) + a(12)*2*p(2) + $ \\
  &\hspace{.5cm} &\hspace{.5cm}\hspace{1cm} $a(10)*p(1)$ \\
$p(1) = m(1) + m(2)$ &\hspace{.5cm} $dp(1) = dm(1) + dm(2)$ &\hspace{.5cm} $a(7) = c(1) + a(17)*p(5) + a(14)*p(3) + $\\
 & & \hspace{.5cm}\hspace{1cm} $ a(11)*2*p(1) + a(10)*p(2)$ \\
$p(0) = m(0)$ &\hspace{.5cm} $dp(0) = 0$ &\hspace{.5cm} $a(6)=pp(0) = c(0)$ \\
$m(4) = m(1)*m(2)$ &\hspace{.5cm} $dm(4) = dm(1)*m(2) + m(1)*dm(2)$ &\hspace{.5cm} $a(5) = a(9)$ \\
$m(3) = \tau(1)$ &\hspace{.5cm} $dm(3) = d\tau(1)$ &\hspace{.5cm} $a(4) = a(8)$ \\
$m(2) = \tau(2)$ &\hspace{.5cm} $dm(2) = d\tau(2)$ &\hspace{.5cm} $a(3) = a(7) + a(5)*m(1)$ \\
$m(1) = \tau(3)$ &\hspace{.5cm} $dm(1) = d\tau(3)$ &\hspace{.5cm} $a(2) = a(7) + a(5)*m(2)$ \\ 
$m(0) = 1$ &\hspace{.5cm} $dm(0) = 0$ &\hspace{.5cm} $a(1) = a(6)$ \\
\noalign{\smallskip}\hline
\hline\noalign{\smallskip}
\end{tabular*}
\end{scriptsize}
\end{table}

We next consider the example of a single water molecule.  The permutational symmetry is 21, meaning that the two hydrogens permute with one another and the oxygen does not permute.  We'll use a maximum polynomial of order 3.  The $MSA$ output is listed in the first column of Table SM-II, while the derivative equations are listed in the second column.  Here is a slightly different way of thinking about the adjoints in the third column.  The operative equation from the main text is
\begin{equation}
a_j(t_j)= c_i \delta_{t,p} +\sum_{i=j+1}^{i_{max}} a_i \frac{\partial{t_i}}{\partial{t_j}}.
\label{eq: adjointrule}
\end{equation}
Note that the partial derivative in the second term of this equation is that of $t_i$ with respect to $t_j$, where $i>j$. Consider first adjoints whose conjugate variables are among the $p$.  The first term contributes a $c_i$ for these. In order for there to be second-term contributions, one or more of the $p_i$ must contain a $dp_j$ term, so for any adjoint we look at the $dp_i$ definitions in the center column to see if any of the right-hand sides contains the derivative of the conjugate variable we are considering, $dp_j$.  For example, $a(17)$ does not have a second term because the rhs of the $dp(12)$ equation does not contain $dp(11)$. Similarly $a(15)$, $a(16)$ and $a(17)$ do not have second terms because the rhs of the $dp(12)$, $dp(11)$, and $dp(10)$ definitions do not contain either $dp(11)$, $dp(10)$ or $dp(9)$. However, for $a(14)$, the rhs of the $dp(11)$ definition does contain $dp(8)$, and the derivative of $dp(11)$ with respect to $dp(8)$ is (-1).  Thus, the second term will be $a(17)$, the adjoint of $p(11)$, times the derivative, (-1). The remainder of the adjoints corresponding to $p$ conjugate variables can be likewise evaluated.  Then we come to the adjoints with $m$ conjugate variables. The adjoint $a(5)$ has conjugate variable $m(4)$ which appears in the definition of $dp(3)$, whose adjoint is $a(9)$. The derivative is 1, so $a(5)=a(9)$. Similarly, the adjoint $a(4)$ has conjugate variable $m(3)$, which appears in the definition of $dp(2)$, chose adjoint is $a(8)$; the derivative is again 1.  Thus $a(4)=a(8)$. The adjoints $a(3)$ and $a(2)$ each have two terms.

We now focus on $a(4)$, $a(3)$, and $a(2)$, since their conjugate variables are, respectively, $dm(3)=d\tau(1)$, $dm(2)=d\tau(2)$, and $dm(1)=d\tau(3)$.  Recalling that the adjoint is the partial derivative of $V$ with respect to the adjoint's conjugate variable, we see that these three adjoints give us, respectively $\frac{\partial{V}}{\partial{\tau(1)}}$, $\frac{\partial{V}}{\partial{\tau(2)}}$, and $\frac{\partial{V}}{\partial{\tau(3)}}$. As in the case of the diatomic molecule, we now use a chain rule to get the desired partial derivatives of $V$ with respect to the Cartesian coordinates. In this case, however, we have three variables, so $\frac{\partial{V}}{\partial{\alpha_n}}=\sum_{m=1}^3\frac{\partial{V}}{\partial{\tau(m))}}\frac{\partial{\tau(m)}}{\partial{\alpha_n)}}=\sum_{m=1}^3\frac{\partial{V}}{\partial{\tau(m))}}\frac{\partial{\tau(m)}}{\partial{r(m)}}\frac{\partial{r(m)}}{\partial{\alpha_n}}$. Two terms contribute to each partial derivative of $V$ with respect to any $\alpha_n$. The results are:
\begin{equation}
\begin{split}
   \frac{\partial{V}}{\partial{x_1}} = &+ a(4))*(-m(3)/a)*\frac{\partial{r(1)}}{\partial{x_1}} + a(3)*(-m(2)/a)*\frac{\partial{r(2)}}{\partial{x_1}} \\
   \frac{\partial{V}}{\partial{y_1}} = & +a(4)*(-m(3)/a)*\frac{\partial{r(1)}}{\partial{y_1}} + a(3)*(-m(2)/a)*\frac{\partial{r(2)}}{\partial{y_1}} \\
   \frac{\partial{V}}{\partial{z_1}} =&  +a(4)*(-m(3)/a)*\frac{\partial{r(1)}}{\partial{z_1}} + a(3)*(-m(2)/a)*\frac{\partial{r(2)}}{\partial{z_1}}\\
   \frac{\partial{V}}{\partial{x_2}} =&  + a(4)*(-m(3)/a)*\frac{\partial{r(1)}}{\partial{x_2}} + a(2)*(-m(1)/a)*\frac{\partial{r(3)}}{\partial{x_2}}\\
 \frac{\partial{V}}{\partial{y_2}} =&  +a(4)*(-m(3)/a)*\frac{\partial{r(1)}}{\partial{y_2}} + a(2)*(-m(1)/a)*\frac{\partial{r(3)}}{\partial{y_2}}\\
    \frac{\partial{V}}{\partial{z_2}} =&  +a(4)*(-m(3)/a)*\frac{\partial{r(1)}}{\partial{z_2}} + a(2)*(-m(1)/a)*\frac{\partial{r(3)}}{\partial{z_2}}\\
    \frac{\partial{V}}{\partial{x_3}} =&  +a(3)*(-m(2)/a)*\frac{\partial{r(2)}}{\partial{x_3}} + a(2)*(-m(1)/a)*\frac{\partial{r(3)}}{\partial{x_3}}\\
    \frac{\partial{V}}{\partial{y_3}} =&  +a(3)*(-m(2)/a)*\frac{\partial{r(2)}}{\partial{y_3}} + a(2)*(-m(1)/a)*\frac{\partial{r(3)}}{\partial{y_3}}\\
    \frac{\partial{V}}{\partial{z_3}} =&  +a(3)*(-m(2)/a)*\frac{\partial{r(2)}}{\partial{z_3}} + a(2)*(-m(1)/a)*\frac{\partial{r(3)}}{\partial{z_3}}\\
\end{split}
\end{equation}

\section*{Reverse Derivative Mathematica Code for an Adjoint}

\begin{table}[ht]
\caption*{Table SM-III. Mathematica Code for an Adjoint}
\centering
\begin{scriptsize}
\begin{tabular*}{\textwidth}{ l  }
\hline
\hline\noalign{\smallskip}

\texttt{GetAdjoint[jval\_,pliststr\_,mliststr\_,qliststr\_,mpqtxt\_,mpqtxttab\_,
natoms\_,npoly\_,nq\_,nmono\_,nvar\_]:=Module[}\\
\texttt{\{fouttxtadd,jend,skipm,return,ltri,ltrj,dolist,eqpos,numi,numj,
lookfor,lasti\}, (* local variables *)}\\
\texttt{(* 
\textcolor{blue}{jval} is the index of mpqtxt or mpqtttab for which you are trying to find the adjoint. }\\
\texttt{\textcolor{blue}{pliststr, mliststr}, and \textcolor{blue}{qliststr} are text strings of the list of p's, m's, and q's (if present).}\\
\texttt{They are created in the calling function by these commands: }\\
\hspace{0.5cm}\texttt{pliststr="";Do[(pliststr=pliststr\textless\textgreater''p''\textless\textgreater ToString[i-1]\textless\textgreater",";),\{i,1,npoly+1\}]; }\\
\texttt{\hspace{0.5cm}qliststr=""; If[nq\textgreater0,Do[(qliststr=qliststr\textless\textgreater"q"\textless\textgreater ToString[i]\textless\textgreater",";),\{i,1,nq\}];];}\\
\texttt{\hspace{0.5cm}mliststr="";
Do[(
mliststr=mliststr\textless\textgreater''m''\textless\textgreater ToString[i-1]\textless\textgreater",";
),\{i,1,nmono+1\}];}\\
\texttt{\textcolor{blue}{mpqtxt} is the Mathematica code for all m's, p's and q's (if present); its format
looks something like this: }\\
\texttt{
\{"m0=1.0D0","m1=x3;","m2=x2;","m3=x1;","m4=m1*m2;","p0=m0; ","p1=m1+m2;","p2=m3;", }\\ \texttt{"p3=m4;","p4=p2*p1;""p5=p1*p1-p3-p3;","p6=p2*p2;",
"p7=p2*p3;","p8=p3*p1;",
"p9=p2*p5;","p10=p2*p4;", }\\ \texttt{"p11=p1*p5-p8;","p12=p2*p6;"\} }\\
\texttt{\textcolor{blue}{mpqtxttab} is the Fortran code for m's, p's and q's in a Table form of text entries, eg.
each term is like }\\ 
\texttt{\{"m(99)","=","m(8)","*","m(24)"\}. }\\
\texttt{\textcolor{blue}{natoms, npoly, nq, nmono,} and \textcolor{blue}{nvar} are, respectively the numbers of atoms (in the
parent),  the numbers of p }\\ \texttt{polynomials, 
the number of q polynomials, the number of 
monomials, and the number of variables (x). }\\

\texttt{\textcolor{blue}{fouttxtadd} is the Fortran code output, in text format. 
The adjoints have the format of }\\ \texttt{pp(i), qp(j), mp(k), 
xxp(n) for the p's, q's, m's, and x's, respectively.  
*) }\\
\\
\texttt{\textcolor{red}{(* get starting part of fouttxtadd, i.e., add c(i) if
the conjugate variable of the adjoint is a p *)} }\\
\texttt{ltrj=StringTake[mpqtxt[[jval]],1]; }\\
\texttt{eqpos=StringPosition[mpqtxt[[jval]],"="][[1,1]]; }\\
\texttt{numj=ToExpression[StringDrop[StringDrop[mpqtxt[[jval]],
\{eqpos,StringLength[mpqtxt[[jval]]]\}],1]]; }\\
\texttt{fouttxtadd=ltrj\textless\textgreater "p"\textless\textgreater ToString[numj]\textless\textgreater" = "; }\\
\texttt{If[ltrj=="p",fouttxtadd=fouttxtadd\textless\textgreater"+ c"\textless\textgreater ToString[numj];]; }\\
\\
\texttt{\textcolor{red}{(* accumulate in dolist the indicies of all mpqtxttab elements that have 
ltrj\textless\textgreater "("\textless\textgreater ToString[jval]\textless\textgreater ")"} }\\ \texttt{on the rhs.  These are the ones whose partial
derivatives will need to be evaluatied *) }\\
\texttt{dolist=\{\}; }\\
\texttt{lookfor=ltrj\textless\textgreater"("\textless\textgreater ToString[numj]\textless\textgreater")"; }\\
\texttt{lasti=Position[mpqtxttab[[All,1]],lookfor][[1,1]]; }\\
\texttt{Do[( }\\
\texttt{\hspace{0.5 cm}If[MemberQ[Drop[mpqtxttab[[i]],2],lookfor],
dolist=Append[dolist,i];
]; }\\
\texttt{),\{i,Length[mpqtxttab],lasti,-1\}]; }\\
\texttt{If[dolist==\{\}, 
If[StringTake[fouttxtadd,-2]=="= ",fouttxtadd="";];
Goto[return];
]; }\\
\\
\\
\\
\\
(Table continued on next page)\\
\noalign{\smallskip}\hline
\hline\noalign{\smallskip}
\end{tabular*}
\end{scriptsize}

\end{table}

\begin{table}[ht]
\caption*{Table SM-III (continued). Mathenatica Code for and Adjoint}
\centering
\begin{scriptsize}
\begin{tabular*}{\textwidth}{ l  }
\hline
\hline\noalign{\smallskip}
\texttt{\textcolor{red}{(* perform do loop over the members of dolist*)} }\\
\texttt{Do[( }\\
\texttt{idyn=i; (* global  for dynamic display of progress*) }\\
\texttt{Quiet[
ToExpression["Clear["\textless\textgreater pliststr \textless\textgreater"];"]; (* these much be cleared  *) }\\
\texttt{ToExpression["Clear["\textless\textgreater mliststr \textless\textgreater"];"]; (* so that derivatives will be *) }\\
\texttt{If[nq!=0,ToExpression["Clear["\textless\textgreater qliststr\textless\textgreater"];"];]; (* symbolic  *)
]; }\\
\texttt{ltri=StringTake[mpqtxt[[i]],1]; }\\
\texttt{eqpos=StringPosition[mpqtxt[[i]],"="][[1,1]]; }\\
\texttt{numi=ToExpression[StringDrop[StringDrop[mpqtxt[[i]],
{eqpos,StringLength[mpqtxt[[i]]]}],1]]; }\\
\texttt{ToExpression[mpqtxt[[i]]]; }\\
\texttt{(* this statement adds the new text in FortranForm after
solving symbolically for the derivative: *) }\\
\texttt{\hspace{0.5cm}fouttxtadd=fouttxtadd\textless\textgreater" + "\textless\textgreater ltri\textless\textgreater"p"\textless\textgreater ToString[numi]\textless\textgreater"*"\textless\textgreater }\\
\texttt{\hspace{0.5cm}\hspace{0.5cm}ToString[FortranForm[D[ToExpression[ltri\textless\textgreater ToString[numi]], }\\
\texttt{\hspace{0.5cm}\hspace{0.5cm}ToExpression[ltrj<>ToString[numj]]]]]; }\\
\texttt{),\{i,dolist\}];  (* end of do loop over dolist *) }\\
\texttt{(* if no rhs, then no new text, thus: *) }\\
\texttt{If[StringTake[fouttxtadd,-2]=="= ",fouttxtadd="";]; }\\
\\
\texttt{\textcolor{red}{(* finish up *)} }\\
\texttt{Label[return]; }\\
\texttt{
If[fouttxtadd!="",fouttxtadd=fouttxtadd \textless\textgreater "(* CR here *) }\\\texttt{n";]; 
}\\

\texttt{ fouttxtadd (* output *) }\\
\texttt{]; }\\
\noalign{\smallskip}\hline
\hline\noalign{\smallskip}
\end{tabular*}
\end{scriptsize}

\end{table}

We list in Table SM-III the Mathematica Code used to find the $j^{th}$ adjoint of a computational plan such as the ones produced by $MSA$ software and, simultaneously, to output Fortran code that can be used to implement it. Comments are enclosed in ``(* ...  *)''. 
\clearpage
\vspace{2.5cm}

\section*{Brief Description of Mathematica Programs}
The current versions of Mathematica programs are here briefly described, with emphasis on the work path for producing purification, compaction, and the appending of derivative programs.  In general, the first step is to use the permutational symmetry and polynomial order to generate $MSA$ Fortran output.\cite{Xie-nma,msachen,Xie10,persp9}  We then convert this output to a ``standard'' Fortran form using a program that allows for fragmentation\cite{NandiBowman2019,conte20_efficientPIP} but can also be used (specifying the parent molecule as a single fragment) to produce the standard form.  The output is a Fortran program. The second step is to use this Fortran program as an input to the Mathematica purification routine, which sorts the polynomials into those that have the correct limiting behavior at large distances and those that do not.  As an option, this program will also perform the ``compaction'' step, which deletes those polynomials and monomials that are no longer needed as a result of the purification.  A further option allows one to include various forms of derivative programs, including the ``normal'' derivative routine, the ``fast (forward)'' derivative routine, or the reverse derivative routine. These derivative routines can be left out of the purified/compacted Fortran output and appended separately in a subsequent Mathematica step.  This Mathematica program also allows the derivatives to be generated and appended to programs that do not need purification.  A further option of the Mathematica purification program is to increase the number of polynomials\cite{conte20_efficientPIP} and coefficients by generation new PIPs via multiplication of the purified ones and selection of those that have the largest value when calculated using the maximum values of the transformed internuclear distances that occur in the data set. A similar add-on step allows ``pruning'' of polynomials, i.e., elimination of the least important polynomials, those that have the smallest values of those calculated in the same manner.\cite{conte20_efficientPIP}  Collectively, these Mathematica-based tools allow one to generate an efficient basis of PIPs, whose number can be increased or decreased, whose long-range limiting behaviour can be controlled, and whose analytical derivative method can be selected.  

\section*{A New Ethanol PES}

To examine the fidelity of this new PES, we performed geometry optimization and normal mode analysis. The agreement with the direct \textit{ab initio} one is excellent. We get the energy of the  global minimum geometry as -154.9995845 Hartree, whereas the direct calculation gives -154.9995972 Hartree. Comparison of harmonic frequencies with their corresponding \textit{ab initio} ones given in Table. SM-IV. It is seen that the agreement with the direct B3LYP/6-311+G(d,p) frequencies is very good; the maximum error is 11 cm$^{-1}$, but most of the frequencies are within couple of cm$^{-1}$ of the \textit{ab initio} ones.

\begin{table}[htbp!]
\caption*{Table SM-IV. Comparison of harmonic frequencies (in cm$^{-1}$) between PES and the corresponding \textit{ab initio} (B3LYP/6-311+G(d,p)) ones of Ethanol.}
\label{tab:PES_Freq}

\begin{threeparttable}
	\begin{tabular*}{\columnwidth}{@{\extracolsep{\fill}}cccccccc}
	\hline
	\hline\noalign{\smallskip}
     Mode & PES & \textit{ab initio} & Diff. & Mode & PES & \textit{ab initio} & Diff. \\
	\noalign{\smallskip}\hline\noalign{\smallskip}
     1  & 237  & 229  &  8  & 12 & 1446 & 1444 &  2 \\
     2  & 268  & 269  & -1  & 13 & 1482 & 1481 &  1 \\
     3  & 419  & 416  &  3  & 14 & 1498 & 1498 &  0 \\
     4  & 821  & 821  &  0  & 15 & 1522 & 1526 & -4 \\
     5  & 896  & 895  &  1  & 16 & 2979 & 2975 &  4 \\
     6  & 1036 & 1028 &  8  & 17 & 3005 & 3001 &  4 \\
     7  & 1094 & 1093 &  1  & 18 & 3031 & 3030 &  1 \\
     8  & 1175 & 1175 &  0  & 19 & 3097 & 3096 &  1 \\
     9  & 1267 & 1256 &  11 & 20 & 3105 & 3103 &  2 \\
     10 & 1298 & 1299 & -1  & 21 & 3841 & 3840 &  1 \\
     11 & 1401 & 1403 & -2  & --- & --- & --- & --- \\
	\noalign{\smallskip}\hline
	\hline
   
	\end{tabular*}

\end{threeparttable}
\end{table}
\clearpage
\bibliography{refs.bib}